\pdfoutput=1

\documentclass[11pt,twoside,a4paper,cmspaper,final,collab]{cms-tdr}
\def\svnVersion{451906P}\def\cmsCernNoTag{CERN-EP-2017-170}\def\cmsCernDate{\today}\def\cmsMessage{Published in the Journal of High Energy Physics as \href{http://dx.doi.org/10.1007/s13130-018-7845-2}{\doi{10.1007/s13130-018-7845-2.}}}

\begin{document}\cmsNoteHeader{SUS-16-034}

\hyphenation{had-ron-i-za-tion}
\hyphenation{cal-or-i-me-ter}
\hyphenation{de-vices}
\RCS$Revision: 442203 $
\RCS$HeadURL: svn+ssh://svn.cern.ch/reps/tdr2/papers/SUS-16-034/trunk/SUS-16-034.tex $
\RCS$Id: SUS-16-034.tex 442203 2018-01-22 12:15:34Z olivito $

\newcommand{\nb}{\ensuremath{N_{\text{b-jets}}}\xspace}
\newcommand{\lint}{35.9\fbinv}
\newcommand{\mll}{\ensuremath{m_{\ell\ell}}\xspace}
\newcommand{\mlb}{\ensuremath{\Sigma m_{\ell\PQb}}\xspace}
\newcommand{\mjj}{\ensuremath{m_{\text{jj}}}\xspace}
\newcommand{\mbb}{\ensuremath{m_{\PQb\PQb}}\xspace}
\newcommand{\secondchi}{\PSGczDt}
\newcommand{\firstchi}{\PSGczDo}
\newcommand{\firstcharg}{\PSGcpmDo}
\newcommand{\gravitino}{\sGra}
\newcommand{\gluino}{\PSg}
\newcommand{\slep}{\ensuremath{\widetilde{\ell}}\xspace}
\newcommand{\sbottom}{\PSQb}
\newcommand{\MuMu}{\ensuremath{\Pgm^\pm\Pgm^\mp}\xspace}
\newcommand{\ElEl}{\ensuremath{\Pe^\pm\Pe^\mp}\xspace}
\newcommand{\EM}{\ensuremath{\Pe^\pm\Pgm^\mp}\xspace}
\newcommand{\rmue}{\ensuremath{r_{\Pgm/\Pe}}\xspace}
\newcommand{\rmuec}{\ensuremath{C}\xspace}
\newcommand{\Rsfof}{\ensuremath{R_{\text{SF/DF}}}\xspace}
\newcommand{\RT}{\ensuremath{R_{\text{T}}}\xspace}
\newcommand{\rinout}{\ensuremath{r_{\text{out/in}}}\xspace}
\newcommand{\PV}{\ensuremath{\cmsSymbolFace{V}\xspace}}
\newcommand{\vz}{\ensuremath{\PV\PZ}\xspace}
\newcommand{\vv}{\ensuremath{\PV\PV}\xspace}
\newcommand{\vvv}{\ensuremath{\PV\PV\PV}\xspace}
\newcommand{\znu}{\ensuremath{\PZ{+}\nu}\xspace}
\newcommand{\dyjets}{\ensuremath{\text{DY}{+}\text{jets}}\xspace}
\newcommand{\gjets}{\ensuremath{\cPgg{+}\text{jets}}\xspace}
\newcommand{\ttz}{\ensuremath{\ttbar\PZ}\xspace}
\newcommand{\ttv}{\ensuremath{\ttbar\PV}\xspace}
\newcommand{\njets}{\ensuremath{N_{\text{jets}}}\xspace}
\newcommand{\mt}{\ensuremath{M_{\text{T}}}\xspace}
\newcommand{\mttwo}{\ensuremath{M_{\text{T2}}}\xspace}
\newcommand{\mttwol}{\ensuremath{M_{\text{T2}}(\ell\ell)}\xspace}
\newcommand{\mttwolb}{\ensuremath{M_{\text{T2}}(\ell \PQb \ell \PQb)}\xspace}
\newcommand{\PZg}{\ensuremath{\PZ/\cPgg^{*}}\xspace}
\providecommand{\NA}{\ensuremath{\text{---}}\xspace}
\newcommand{\x}{\ensuremath{\phantom{0}}}
\newcommand{\y}{\ensuremath{\phantom{.}}}

\cmsNoteHeader{SUS-16-034}
\title{Search for new phenomena in final states with two opposite-charge, same-flavor leptons, jets, and missing transverse momentum in pp collisions at \texorpdfstring{$\sqrt{s}=13\TeV$}{sqrt(s) = 13 TeV}}

\date{\today}

\abstract{
Search results are presented for physics beyond the standard model in final states with two opposite-charge, same-flavor leptons, jets, and missing transverse momentum. The data sample corresponds to an integrated luminosity of  35.9\fbinv of proton-proton collisions at $\sqrt{s}=13\TeV$ collected with the CMS detector at the LHC in 2016. The analysis uses the invariant mass of the lepton pair, searching for a kinematic edge or a resonant-like excess compatible with the Z boson mass. The search for a kinematic edge targets production of particles sensitive to the strong force, while the resonance search targets both strongly and electroweakly produced new physics. The observed yields are consistent with the expectations from the standard model, and the results are interpreted in the context of simplified models of supersymmetry.  In a gauge mediated supersymmetry breaking (GMSB) model of gluino pair production with decay chains including Z bosons, gluino masses up to 1500--1770\GeV are excluded at the 95\% confidence level depending on the lightest neutralino mass. In a model of electroweak chargino-neutralino production, chargino masses as high as 610\GeV are excluded when the lightest neutralino is massless.  In GMSB models of electroweak neutralino-neutralino production, neutralino masses up to 500--650\GeV are excluded depending on the decay mode assumed. Finally, in a model with bottom squark pair production and decay chains resulting in a kinematic edge in the dilepton invariant mass distribution, bottom squark masses up to 980--1200\GeV are excluded depending on the mass of the next-to-lightest neutralino.
}

\hypersetup{%
pdfauthor={CMS Collaboration},%
pdftitle={Search for new phenomena in final states with two opposite-sign, same-flavor leptons, jets, and missing transverse momentum in pp collisions at sqrt(s) =13 TeV},%
pdfsubject={CMS},%
pdfkeywords={CMS, physics, SUSY}}

\maketitle

\section{Introduction}
\label{sec:introduction}

Supersymmetry (SUSY)~\cite{Ramond:1971gb,Golfand:1971iw,Neveu:1971rx,Volkov:1972jx,Wess:1973kz,Wess:1974tw,Fayet:1974pd,Nilles:1983ge}
is a well-studied extension of the standard model (SM) and assumes a new fundamental
symmetry that assigns a fermion (boson) to each SM boson (fermion). Supersymmetry resolves the hierarchy problem
by stabilizing the Higgs boson (H) mass via additional quantum loop corrections from the top quark superpartner (top squark),
which compensate for the large correction due to the top quark.
If $R$-parity~\cite{Farrar:1978xj} is conserved, the lightest SUSY particle (LSP)
predicted by the theory is stable and potentially massive, providing
a candidate for the observed dark matter. Many SUSY models also lead to the unification of the electroweak (EW) and strong forces at high energies~\cite{unification,unification2}.

This paper presents a search for signatures of new physics in events with two opposite-charge, same-flavor (OCSF) leptons (electrons or muons),
jets, and missing transverse momentum. Interpretations of the search results are given in terms of simplified supersymmetric model spectra.
The data set of proton-proton collisions used for this search was collected in 2016 with the CMS detector at the CERN LHC
at a center-of-mass energy of $\sqrt{s} = 13\TeV$ and corresponds to an integrated luminosity of \lint.
Final states including an OCSF dilepton pair can occur in SUSY models via the decay of the superpartner of the SM neutral gauge bosons,
the neutralino, when a heavier neutralino decays to a lighter neutralino LSP,
or when the lightest neutralino is the next-to-lightest SUSY particle decaying to a gravitino LSP.
Depending on the model parameters, the neutralino can decay into the LSP and either an on-shell \PZ boson or a
virtual $\PZ/\gamma^{*}$ boson, with the boson decaying to two charged leptons.
The neutralino can also possibly decay into a lepton and its supersymmetric partner (slepton),
the latter decaying into another lepton and the LSP. Decays involving an on-shell \PZ boson are expected to produce an excess of events
in which the dilepton invariant mass is compatible with the \PZ boson mass, referred to as the "on-\PZ signature",
while decays involving off-shell \PZ bosons or sleptons are expected to produce a characteristic edge shape in the invariant mass distribution of
the dilepton system (\mll)~\cite{Hinchliffe:1996iu}, denoted as the "edge signature".

This search targets both the on-\PZ and edge signatures.
For the on-\PZ signature, search regions are optimized separately depending on whether we target strong or EW SUSY production.
In the case of strong production, the neutralino is part of a decay chain starting from a gluino or squark, while in the
EW case, it is directly produced.
The search for a kinematic edge is only performed under the assumption of strong SUSY production.

Searches for SUSY in these final states were performed previously by the CMS~\cite{CMS:edge,CMS:Zedge2015,OSpaperCMS7TeV,OSpaperCMS2011,2012ewk,2012ewkhiggs} and ATLAS~\cite{ATLAS:edge,ATLASewk8tev,ATLASOS13tev} Collaborations.
The CMS Collaboration reported the presence of an excess with an edge shape located at $\mll=78.7\pm1.4\GeV$ and with a local significance of 2.4 standard deviations (s.d.) in the data set collected at a center-of-mass-energy of $\sqrt{s} = 8\TeV$~\cite{CMS:edge}.
The ATLAS Collaboration did not confirm this excess in its $\sqrt{s} = 8\TeV$ dataset, but reported a resonant-like excess of events compatible with the \PZ boson mass and with a local significance of 3.0 s.d.~\cite{ATLAS:edge}. Neither of these excesses were confirmed in the data sets collected at a center-of-mass-energy of $\sqrt{s} = 13\TeV$ during 2015 by the CMS Collaboration~\cite{CMS:Zedge2015} and during 2015 and the first half of 2016 by the ATLAS Collaboration~\cite{ATLASOS13tev}.
\section{Signal models}
\label{sec:signalmodels}

The results of this search are interpreted in the context of various
simplified models of SUSY~\cite{bib-sms-1,bib-sms-2,bib-sms-3,bib-sms-4,Chatrchyan:2013sza},
as described below.
In all models, the \PW, \PZ, and Higgs bosons are assumed to decay according to their SM branching fractions.

\begin{figure}[tb]
  \centering
    \includegraphics[width=0.4\textwidth]{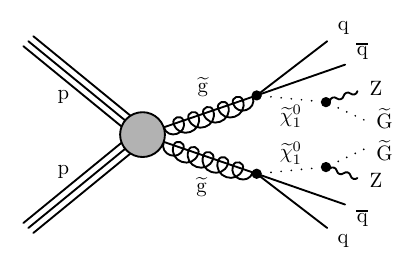}
    \includegraphics[width=0.4\textwidth]{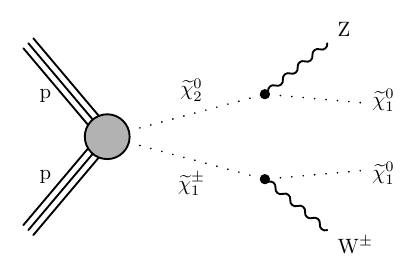}
    \includegraphics[width=0.4\textwidth]{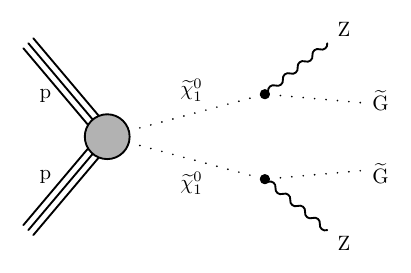}
    \includegraphics[width=0.4\textwidth]{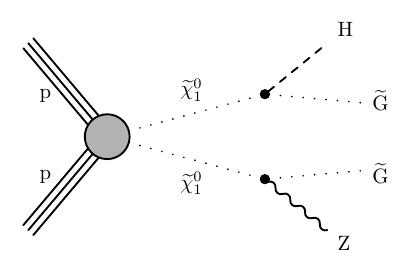}
  \caption{
    \label{sig:feynmanOnZ}
    Diagrams for models with decays containing at least one dilepton pair stemming from an on-shell \PZ\ boson decay studied in this analysis.
    The model targeted by the strong-production search is shown in the upper left.
    The three other diagrams correspond to EW production of chargino-neutralino or neutralino-neutralino pairs.
    All the diagrams containing a gravitino (\gravitino) represent gauge-mediated SUSY breaking (GMSB) models.
  }
\end{figure}

This search is designed to be sensitive to both strong and EW SUSY production
leading to the on-\PZ signature.
Most of the simplified models used for interpretation of the on-\PZ results represent gauge-mediated supersymmetry breaking
(GMSB) models~\cite{Matchev:1999ft,Meade:2009qv,Ruderman}.
The first of these GMSB models assumes strong production of a pair of gluinos (\gluino) that each decays into a pair of
quarks (\PQu, \PQd, \PQs, \PQc, or \PQb) and the lightest neutralino, \firstchi.
The \firstchi\ in turn decays into a massless gravitino (\gravitino) and an on-shell \PZ boson.
The decay chain corresponding to this gluino GMSB model is shown in Fig.~\ref{sig:feynmanOnZ} (upper left).

The three other models used for the on-\PZ signature assume EW production.
The upper right diagram in Fig.~\ref{sig:feynmanOnZ} corresponds to chargino-neutralino (\firstcharg-\secondchi) production,
with \firstcharg decaying to a \PW\ boson and the LSP, \firstchi,
while the next-to-lightest neutralino, \secondchi, decays to a \PZ boson and \firstchi.
The production cross sections for this model are computed in a limit of mass-degenerate wino \firstcharg\ and \secondchi,
and light bino \firstchi, with all the other sparticles assumed to be heavy and decoupled.
Gauge mediated supersymmetry breaking is not assumed for this $\PW\PZ$ model, and the \firstchi is allowed to be massive.

The remaining two models considered assume the production of neutralino-neutralino (\firstchi-\firstchi) pairs in GMSB.
For bino- or wino-like neutralinos, the neutralino pair production cross section is very small, and thus 
we consider a specific GMSB model with mass-degenerate higgsinos \firstcharg, \secondchi, and \firstchi as the next-to-lightest SUSY particles 
and a massless gravitino as the LSP~\cite{Matchev:1999ft,Meade:2009qv,Ruderman}.
In the production of any two of these, \firstcharg\ or \secondchi\
decays immediately to \firstchi\ and low-momentum particles that do not impact the analysis,
effectively yielding pair production of $\firstchi\firstchi$.
Intermediate production of either \firstcharg\ or \secondchi\ is therefore not explicitly shown
in the lower two diagrams of Fig.~\ref{sig:feynmanOnZ} representing these models.
In the first model (lower left of Fig.~\ref{sig:feynmanOnZ}),
the only allowed decay of the lightest neutralino is to a \PZ boson and a massless gravitino.
In the other model (lower two diagrams of Fig.~\ref{sig:feynmanOnZ}),
the lightest neutralino is allowed to decay to a gravitino and either a \PZ boson or an SM-like Higgs boson,
with a 50\% branching fraction to each decay channel.
The cross sections for higgsino pair production are computed in a limit of mass-degenerate higgsino states \secondchi,
\firstcharg, and \firstchi, with all the other sparticles assumed to be heavy and decoupled.
Following the convention of real mixing matrices and signed neutralino masses~\cite{Skands:2003cj},
we set the sign of the mass of \firstchi (\secondchi) to $+1$ ($-1$).
The lightest two neutralino states are defined as symmetric (anti-symmetric) combinations
of higgsino states by setting the product of the elements $N_{i3}$ and $N_{i4}$
of the neutralino mixing matrix $N$ to $+0.5$ ($-0.5$) for $i = 1$ ($2$).
The elements $U_{12}$ and $V_{12}$ of the chargino mixing matrices $U$ and $V$ are set to 1.

\begin{figure}[tb]
  \centering
    \includegraphics[width=0.4\textwidth]{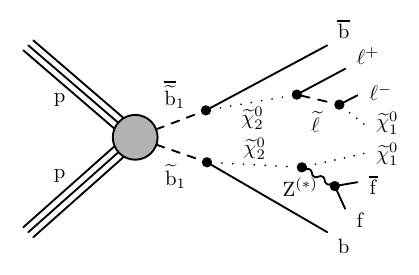}
  \caption{
    \label{sig:feynmanEdge}
    Diagram showing a possible decay chain in the slepton edge model.
    Bottom squarks are pair produced with subsequent decays that frequently contain dilepton pairs.
    This model features a characteristic edge in the \mll\ spectrum given approximately
    by the mass difference between the \secondchi\ and \firstchi\ particles.
  }
\end{figure}

The signal model for the edge search, referred to as the slepton edge model, assumes the production of a pair of bottom squarks (\sbottom),
the superpartner of the bottom quark, where each
decays to \secondchi and a bottom quark. Two decay modes of the \secondchi are considered,
each with a 50\% branching fraction; they are both illustrated in Fig.~\ref{sig:feynmanEdge}.
In the first mode, the \secondchi decays to a \PZ boson and \firstchi, which is stable.
The \PZ boson can be on- or off-shell, depending
on the mass difference between the neutralinos. The second decay mode features sequential two-body decays with an intermediate
slepton \slep (\PSe,\PSGm): $\secondchi \to\slep \ell \to\ell\ell\firstchi$.
The masses of the sleptons are assumed to be degenerate and
equal to the average of the \secondchi and \firstchi masses. The masses of the \sbottom and \secondchi are free parameters,
while the mass of \firstchi\ is fixed at 100\GeV.
This scheme allows the position of the signal edge to vary in the invariant mass distribution depending on
the mass difference between the \secondchi and \firstchi.
The mass of the \firstchi\ is chosen in such a way that the \secondchi\ mass is always greater by at least 50\GeV,
setting the minimum possible edge position at 50\GeV.

\section{The CMS detector}
\label{sec:cmsdetector}

The central feature of the CMS apparatus is a superconducting solenoid, 13\unit{m} in length and 6\unit{m} in diameter, that provides
an axial magnetic field of 3.8\unit{T}. Within the solenoid volume are various particle detection systems. Charged-particle
trajectories are measured by silicon pixel and strip trackers, covering $0 < \phi < 2\pi$ in azimuth and $\abs{\eta} < 2.5$,
where the pseudorapidity $\eta$ is defined as $-\log [\tan(\theta/2)]$, with $\theta$ being the polar angle of the
trajectory of the particle with respect to the counterclockwise beam direction. A crystal electromagnetic calorimeter (ECAL) and a brass and scintillator
hadron calorimeter (HCAL) surround the tracking volume. The calorimeters provide energy
and direction measurements of electrons and hadronic jets. Muons are detected in gas-ionization detectors embedded in
the steel flux-return yoke outside the solenoid. The detector is nearly hermetic, allowing for momentum balance measurements in the
plane transverse to the beam direction. A two-tier trigger system selects events of interest for physics analysis.
A more detailed description of the CMS detector, together with a definition of the coordinate system used and the relevant kinematic variables,
can be found in Ref.~\cite{Chatrchyan:2008zzk}.

\section{Data sets, triggers, and object selection}
\label{sec:samplesObjects}

This analysis uses data samples of \ElEl and \MuMu events for the signal region (SR) selections
and \EM events for control regions (CRs).
Events are collected with a set of dilepton ($\Pe\Pe$, $\Pgm\Pgm$, or $\Pe\Pgm$) triggers that require the magnitude of the transverse momentum
$\pt > 17$ or 23\GeV for the highest \pt\ lepton,
depending on the data taking period, except for the dimuon trigger where the requirement is always $\pt > 17\GeV$. These triggers
impose loose isolation criteria on the leptons.
For the next-to-highest \pt\ electron (muon), $\pt > 12$ $(8) \GeV$ is required, and electrons (muons) must satisfy $\abs{\eta}< 2.5$ $(2.4)$.
In order to retain high signal efficiency, in particular for highly boosted dilepton systems,
dilepton triggers without an isolation requirement are also used.
These require $\pt > 33\GeV$ for both leptons in the dielectron case and $\pt > 30\GeV$ for both leptons in the electron-muon case.
In the dimuon case, they require either $\pt > 27$ $(8)$ or $\pt > 30$ $(11)\GeV$ for the highest (next-to-highest) \pt\ muon
depending on the data taking period.
The trigger efficiencies are measured in data using events selected by a suite of jet triggers and are found to be 90--96\%.

The particle-flow (PF) event algorithm~\cite{Sirunyan:2017ulk} reconstructs and identifies particle candidates in the event,
referred to as PF objects.
To select collision events we require at least one reconstructed vertex.
The reconstructed vertex with the largest value of summed physics-object
$\pt^2$ is taken to be the primary $\Pp\Pp$ interaction vertex. The physics
objects used for the primary vertex selection are the objects returned by a jet finding
algorithm~\cite{Cacciari:2008gp,FastJet} applied to all charged tracks
associated with the vertex, plus the corresponding associated missing transverse momentum.
The missing transverse momentum vector \ptvecmiss is defined as the projection onto the plane perpendicular to the beam axis
of the negative vector sum of the momenta of all reconstructed PF objects in an event. Its magnitude is referred to as \ptmiss.

Electrons, reconstructed by associating tracks with ECAL clusters, are identified using a multivariate approach based on information on the
cluster shape in the ECAL, track reconstruction quality, and the matching between the track and the ECAL cluster~\cite{Khachatryan:2015hwa}.
Electrons from reconstructed photon conversions are rejected.
Muons are reconstructed from tracks found in the muon system associated with tracks found in the tracker. They are identified
based on the quality of the track fit and the number of associated hits in the tracking detectors. For both lepton flavors, the impact
parameter with respect to the primary vertex
is required to be within 0.5\unit{mm} in the transverse
plane and less than 1\unit{mm} along the beam direction. The lepton isolation variable is defined as the scalar \pt\ sum of all PF objects
in a cone around the lepton (excluding those identified as electrons or muons).
To mitigate the impact of additional $\Pp\Pp$  interactions in the same or nearby bunch crossings (pileup),
only charged PF objects compatible with the primary vertex are included in the sum,
and the average expected pileup contribution is subtracted from the neutral component of the isolation.
The isolation sum is required to be smaller than 10 (20)\% of the lepton \pt for electrons (muons).
The cone size varies with lepton \pt\ and is chosen to be
$\sqrt{(\Delta\phi)^2 + (\Delta \eta)^2} = \DR = 0.2$ for $\pt < 50\GeV$, $\DR = 10\GeV/\pt$ for $50 < \pt < 200\GeV$,
and $\DR = 0.05$ for $\pt > 200\GeV$. This shrinking cone size with increasing \pt\ preserves high efficiency
for leptons from Lorentz-boosted boson decays~\cite{Rehermann:2010vq}.
To identify events with three or more charged leptons, additional leptons beyond the first two are selected with the looser requirement
for the isolation sum to be less than 40\% of the lepton \pt.

Photons are required to pass identification criteria based on the cluster shape in the ECAL
and the fraction of energy deposited in the HCAL~\cite{CMSPhotonID}.
Photons must satisfy $\pt > 25\GeV$, and be within $\abs{\eta}<2.4$, excluding the ``transition region"
of $1.4<\abs{\eta}<1.6$ between the ECAL barrel and endcap.
Photons are required to be isolated from other PF objects within a cone of $\DR = 0.3$.
To ensure the photon is well measured, it is required that $\Delta\phi(\ptvecmiss, \vec{p}^{\cPgg}_\mathrm{T}) > 0.4$.
To distinguish photons from electrons, the photon is rejected if it can be connected to a pattern of hits in the pixel detector
that indicate the presence of a charged particle track.

Isolated, charged-particle tracks identified by the PF algorithm are selected with looser requirements
on a similar set of criteria to the leptons defined above and are used as a veto on the presence of additional charged leptons.
When selecting charged PF objects, a track-based relative isolation is used. The relative track isolation is calculated using all charged
PF objects within a cone $\DR = 0.3$ and longitudinal impact parameter $\abs{\Delta z} < 0.1$\unit{cm} relative to the primary vertex.
Particle-flow objects identified as electrons or muons (charged hadrons)
are required to have $\pt > 5\,(10)\GeV$ and an isolation value less than 20\,(10)\% of the object \pt.

Jets are clustered from PF objects, excluding charged hadrons not associated with the primary vertex,
using the anti-\kt\ clustering algorithm~\cite{Cacciari:2008gp} with a distance parameter of 0.4,
implemented in the \FASTJET\ package~\cite{FastJet,Cacciari:2005hq}.
Jets are required to satisfy $\abs{\eta} < 2.4$ and $\pt > 35\GeV$,
where the \pt is corrected for nonuniform detector response
and multiple collision (pileup) effects~\cite{1748-0221-6-11-P11002,cacciari-2008-659}.
A jet is removed from the event if it lies within $\DR < 0.4$ of any of the selected leptons or the highest \pt\ photon.
The scalar sum of all jet \pt is referred to as \HT.
Corrections to the jet energy are propagated to \ptmiss\
using the procedure developed in Ref.~\cite{1748-0221-6-11-P11002}.
Identification of jets originating from b quarks is performed with the combined secondary vertex (CSVv2) algorithm~\cite{BTV-16-002},
using the medium working point, for which the typical
efficiency for b quarks is around 60--75\% and the mistagging rate for light-flavor jets is around 1.5\%.
Jets with a lower threshold of $\pt > 25\GeV$ are considered, and selected jets are denoted as b-tagged jets.

Events are selected by requiring two OCSF leptons (\ElEl or \MuMu)
with $\pt > 25$ $(20)\GeV$ for the highest (next-to-highest) \pt\ lepton
and $\abs{\eta} < 2.4$ for both leptons. The distance between the leptons must satisfy $\DR > 0.1$
to avoid reconstruction efficiency differences between electrons and muons in events with collinear leptons.
To ensure symmetry in acceptance between electrons and muons, all leptons in the transition region between the barrel and endcap of the ECAL,
$1.4 < \abs{\eta} < 1.6$, are rejected.
A control sample of lepton pairs with opposite charge and different flavor (OCDF),
\EM, is defined using the same lepton selection criteria.
All the parameters above have been chosen in order to maximize the lepton selection efficiency
while keeping the efficiencies similar for electrons and muons.
Photon events are used to predict one of the main backgrounds of this analysis,
and a data control sample is selected as described below in Section~\ref{sub:dybkg}.
To be consistent with the photon \pt\ threshold applied in this control sample,
we require the \pt of the dilepton system to be greater than 25\GeV.

While the main SM backgrounds are estimated using data control samples, simulated events are used to estimate systematic uncertainties and
some SM background components as described below.
Next-to-leading order (NLO) and next-to-NLO cross sections~\cite{Alwall:2014hca,Alioli:2009je,Re:2010bp,Gavin:2010az,Gavin:2012sy,Czakon:2011xx}
are used to normalize the simulated background samples,
while NLO plus next-to-leading-logarithmic (NLL) calculations~\cite{Borschensky:2014cia,Fuks:2012qx,Fuks:2013vua}
are used for the signal samples.
Simulated samples of Drell--Yan (DY) processes and photons produced in association with jets
are generated with the 
{\MADGRAPH{}5\_a\MCATNLO}2.3.3 
event generator~\cite{Alwall:2014hca} to leading order (LO) precision,
with up to four additional partons in the matrix element calculations, using the MLM matching scheme~\cite{Alwall:2007fs}.
Simulated \ttv ($\PV=\PW,\PZ$) and \vvv\ events are produced with the same generator to NLO precision.
Other SM processes, such as \vv, \ttbar, and single top quark production, are simulated using \POWHEG~2.0~\cite{powheg}.
The matrix element calculations performed with these generators are interfaced with \PYTHIA~8.212~\cite{Sjostrand:2007gs} for the
simulation of parton showering and hadronization. The NNPDF3.0 parton distribution functions (PDF)~\cite{Ball:2014uwa} are used for all samples.
The detector response is simulated with a \GEANTfour model~\cite{Geant} of the CMS detector.
The simulation of new-physics signals is performed using the 
{\MADGRAPH{}5\_a\MCATNLO} 
program at LO precision, with up to
two additional partons in the matrix element calculation. Events are then interfaced with \PYTHIA~8.212 for fragmentation and hadronization,
and further processed using the CMS fast simulation package~\cite{fastsim}. Multiple \Pp\Pp~interactions are superimposed on the hard
collision, and the simulated samples are reweighted such that the number of collisions per bunch crossing accurately reflects
the distribution observed in data.
Corrections are applied to the simulated samples to
account for differences between simulation and data in the trigger and reconstruction efficiencies.

\section{Signal regions}
\label{subsec:signalregions}

The selections for all SRs, described below, are summarized in Table~\ref{tab:selections_signalRegions}.

The on-\PZ search regions are designed to achieve low backgrounds from SM processes, while maintaining sensitivity
to a variety of new-physics models, not only the processes described in Section~\ref{sec:signalmodels}.
The dilepton invariant mass is required to be in the range $86 < \mll < 96\GeV$, which is compatible with the \PZ boson mass.
The events must contain at least two jets and satisfy $\ptmiss > 100\GeV$.
The two highest \pt\ jets in the event are required to have a separation in $\phi$ from \ptvecmiss\ of at least 0.4
to reduce backgrounds where the \ptmiss\ in the event comes from jet mismeasurements.
Events containing additional electrons (muons) with $\pt\ > 10\GeV$, $\abs{\eta} < 2.5 (2.4)$,
and passing the looser isolation criteria from Section~\ref{sec:samplesObjects}, are rejected,
as are events containing an isolated, charged PF candidate passing the selections described in that Section.
Multiple on-\PZ SRs are defined using these selection criteria as a baseline:
the first set for the strong production search and two additional regions for EW production searches.

For the on-\PZ strong-production SRs, we make selections requiring a large level of hadronic activity in the event,
which we expect in the decays of strongly coupled new particles.
We define three SR categories: ``SRA'' (2--3~jets), ``SRB'' (4--5~jets), and ``SRC'' ($\geq$6~jets).
These categories are further divided as having either zero or at least one b-tagged jet.
The kinematic variable \mttwo~\cite{MT2variable,MT2variable2} is used to reduce the background from \ttbar\ events.
This variable was first introduced
to measure the mass of pair-produced particles, each decaying to the same
final state, consisting of a visible and an invisible particle.
It is defined using \ptvecmiss and two visible objects (leptons, jets, or combinations thereof) as:
\begin{equation}
\mttwo = \min_{\ptvecmiss{}^{(1)} + \ptvecmiss{}^{(2)} = \ptvecmiss}
  \left[ \max \left( \mt^{(1)} , \mt^{(2)} \right) \right],
\label{eq:MT2}
\end{equation}
where $\ptvecmiss{}^{(i)}$ ($i=1$, 2) are trial vectors
obtained by decomposing \ptvecmiss.  The transverse masses
$\mt^{(i)} = \sqrt{\smash[b]{2 \pt^{\text{vis}} \ptmiss{}^{(i)} [1 - \cos(\Delta\phi)]}}$,
where $\Delta\phi$ is the angle between the
transverse momentum of the visible object and $\ptvecmiss{}^{(i)}$,
are obtained by pairing either of these trial vectors with one of the two visible objects.
The minimization is performed over all trial momenta satisfying the \ptvecmiss\ constraint.
When building \mttwo from the two selected leptons and \ptvecmiss, denoted \mttwol, its distribution exhibits a sharp
decline around the mass of the \PW\ boson for \ttbar\ events and is therefore well suited to suppress this background.
A requirement of $\mttwol > 80\GeV$ (100\GeV for events with at least one b-tagged jet) is imposed in order to suppress \ttbar backgrounds.
Requirements are then placed on \HT depending on the number of jets and on the presence or absence of a b-tagged jet in the event,
indicated by the labels ``b tag'' and ``b veto,'' respectively.
Finally, each SR is divided into multiple bins in \ptmiss, depending on the number of selected jets. 
The precise requirements are summarized in Table~\ref{tab:selections_signalRegions}.

The first EW on-\PZ search region is denoted the ``\vz'' region and is designed to be sensitive to signatures
where a hadronically decaying \PW\ or \PZ boson is produced in conjunction with the leptonically decaying \PZ boson.
In order to reduce the \ttbar background, events with a b-tagged jet are removed, and we require $\mttwol > 80\GeV$.
The two jets in the event that are closest in $\phi$ are then required to have a dijet invariant mass $\mjj < 110\GeV$
to be consistent with the hadronic decay of a \PW\ or \PZ boson.
The SR is then divided into four bins in \ptmiss: 100--150, 150--250, 250--350, and $>350\GeV$.

The second EW-production search region is denoted the ``$\PH\PZ$'' region and is designed to be sensitive to signatures
where a Higgs boson is produced in conjunction with the leptonically decaying \PZ boson.
We target Higgs bosons decaying to \bbbar, due to its dominant branching ratio,
and we therefore require events to have exactly two b-tagged jets with an invariant mass, \mbb, less than 150\GeV.
In order to reduce the \ttbar background, a \mttwo variable is calculated using
two combinations of one lepton and one b-tagged jet as the visible objects.
Each lepton is paired with a b-tagged jet, and all combinations of \mttwo are calculated.
The smallest value of \mttwo is used, denoted \mttwolb.
The distribution of \mttwolb has an endpoint at the top quark mass for \ttbar events, and we require $\mttwolb > 200\GeV$.
The SR is then divided into three bins in \ptmiss: 100--150, 150--250, and $>250\GeV$.
Although the selections for the EW \vz and $\PH\PZ$ regions are mutually exclusive,
they are not necessarily exclusive with respect to the strong-production SR selections.
For interpretations of the analysis results, either the strong-production or the EW regions
are considered depending on the signal model.

The baseline SR in the edge search requires $\mll > 20\GeV$, at least two jets,
$\ptmiss > 150\GeV$, $\mttwol > 80\GeV$, and the two jets with the highest \pt\ to have
a separation in $\phi$ from \ptvecmiss of at least 0.4.
A fit is performed in this baseline region to search for a kinematic edge in the \mll\ spectrum.
A counting experiment is also performed in seven bins of \mll, excluding the range used for the on-\PZ search.
These are summarized in Table~\ref{tab:selections_signalRegions}.

A likelihood discriminant is used to distinguish between events originating from dileptonically decaying top quark pairs and other sources.
The observables used for the likelihood discriminator are \ptmiss, the \pt of the dilepton system, $\abs{\Delta(\phi)}$ between the leptons,
and an observable called \mlb. The latter is the sum of the
invariant masses of the two lepton and b-tagged jet systems, and should have an endpoint at $2\sqrt{M(\PQt)^2-M(\PW)^2}$
for events resulting from top quark pairs. To calculate \mlb, all pairings of a lepton with a jet are considered,
and the pairing with the minimum invariant mass is selected.
This process is repeated for the remaining lepton and jets, and the sum of the invariant masses of the two lepton-jet pairs is defined as \mlb.
If b-tagged jets are present, they are given priority in the calculation of both lepton-jet systems;
i.e., if one or more b-tagged jets are present, \mlb between the leptons and the b-tagged jet(s) is minimized first,
and then the remaining (b-tagged) jets are considered for the minimization of the sum \mlb of the second lepton.
To calculate the likelihood discriminant, the probability density functions of the four observables are determined by fits in the
different-flavor (DF) control sample using the same kinematic requirements as the same-flavor (SF) SR except removing the \mttwol selection.
The respective fit functions are a sum of two exponential functions for \ptmiss,
a second-order polynomial for $\abs{\Delta(\phi)}$, and a Crystal Ball (CB) function~\cite{Crystal} for both the dilepton \pt and \mlb distributions.
A likelihood function is constructed,
and its negative logarithm is taken as the discriminator value.
Two categories of events are defined: ``\ttbar-like,'' with a discriminator value less than 21 and an efficiency of 95\% for \ttbar events,
and ``not-\ttbar-like,'' which is composed of the remainder of the events.

In addition, two aggregate SRs are defined for the edge search,
integrating the mass bins below and above the \PZ boson mass for the not-\ttbar-like category.

\begin{table}[htbp]
\centering
\topcaption{\label{tab:selections_signalRegions} Summary of all SR selections. }
\resizebox{\textwidth}{!}
{ 
\begin{tabular}{llllll}
\hline
\multicolumn{6}{c}{ Strong-production on-\PZ ($86 < \mll < 96\GeV$) signal regions  }  \\
\hline
Region & \njets & \nb & \HT [\GeVns{}]& \mttwol [\GeVns{}]& \ptmiss binning [\GeVns{}]\\
\hline
SRA b veto & 2--3 & $=$0 & $>$500 & $>$80 & 100--150, 150--250, $>$250 \\
SRB b veto & 4--5 & $=$0 & $>$500 & $>$80 & 100--150, 150--250, $>$250 \\
SRC b veto & $\geq$6 & $=$0 & \NA & $>$80 & 100--150, $>$150  \\
SRA b tag  & 2--3 & $\geq$1 & $>$200 & $>$100 & 100--150, 150--250, $>$250 \\
SRB b tag  & 4--5 & $\geq$1 & $>$200 & $>$100 & 100--150, 150--250, $>$250 \\
SRC b tag  & $\geq$6 & $\geq$1 & \NA & $>$100 & 100--150, $>$150  \\

\hline
\multicolumn{6}{c}{Electroweak-production on-\PZ ($86 < \mll < 96\GeV$) signal regions}  \\
\hline
Region & \njets & \nb & Dijet mass [\GeVns{}]& \mttwo [\GeVns{}]& \ptmiss binning [\GeVns{}]\\
\hline
\vz & $\geq$2 & $=$0 & $\mjj < 110$ & $\mttwol > 80$ & 100--150, 150--250, 250--350, $>$350 \\
$\PH\PZ$ & $\geq$2 & $=$2 & $\mbb < 150$ & $\mttwolb > 200$ & 100--150, 150--250, $>$250  \\

\hline
\multicolumn{6}{c}{Edge signal regions}  \\
\hline
Region & \njets & \ptmiss [\GeVns{}]& \mttwol [\GeVns{}]& \ttbar likelihood & \mll  binning [\GeVns{}]\\
\hline
  Edge fit & $\geq$2 & $>$150 & $>$80 & \NA & $>$20 \\

  \multirow{2}{*}{\ttbar-like} & \multirow{2}{*}{$\geq$2} & \multirow{2}{*}{$>$150} & \multirow{2}{*}{$>$80} & \multirow{2}{*}{$<$21} & 20--60, 60--86, 96--150, 150--200, \\
   & & & & & 200--300, 300--400, $>$400 \\

  not-\ttbar-like & $\geq$2 & $>$150 & $>$80 & $>$21 & same as \ttbar-like \\
  aggregate & $\geq$2 & $>$150 & $>$80 & $>$21 & 20--86, $>$96 \\
\hline
\end{tabular}
}
\end{table}

\section{Standard model background predictions}
\label{sec:backgrounds}

The backgrounds from SM processes are divided into three categories. Those that produce DF pairs (\EM)
as often as SF pairs (\MuMu, \ElEl) are referred to as flavor-symmetric (FS) backgrounds. Among them, the
dominant contribution arises from top quark pair production;
subleading contributions are also present from $\PW^+\PW^-$, $\PZg (\to\tau\tau)$,
$\PQt\PW$ single-top quark production, and leptons from hadron decays.
Data samples of DF events are used to predict the SF background.

The remaining background categories contain flavor-correlated sources of lepton production
that only contribute events with OCSF leptons.
The dominant contributions at lower \ptmiss\ are from DY production in association with jets,
where \ptmiss\ arises from mismeasurement of the jet energies.
Data samples of photon events are used to predict this \dyjets background.

The final category comes from events with prompt neutrinos in addition to an OCSF pair from a \PZg boson.
This includes $\PW\PZ$ and $\PZ\PZ$ production and processes with lower cross section such as \ttz among others.
These backgrounds are referred to as ``\znu'' and can be important in the high-\ptmiss signal bins.

\subsection{Flavor-symmetric backgrounds}
\label{sub:fsbkg}

The method of estimating the FS backgrounds relies on the fact that, for such processes, SF and DF events are
produced at the same rate. This allows for prediction of the background yields in the SF sample from those in the DF sample
by application of an appropriate correction factor, which is estimated from CRs in data.
This factor corrects for different flavor-dependent
reconstruction and identification efficiencies and for flavor-dependent trigger efficiencies, which can be different for electrons and muons.

For cases where the DF contribution is of
sufficient statistical power to make an accurate prediction in the SF channel,
a background estimate in the SF channel can therefore be obtained by applying a multiplicative correction
factor, \Rsfof, to the DF channel yield. The correction is determined in two independent ways,
both based purely on control samples in the data.
The two results are then combined using the weighted average
according to their uncertainties to obtain the final factor.
The first approach uses a direct measurement of this correction factor in a data CR independent of the baseline SR, and
the second method involves a factorized approach of measuring the effects of reconstruction, identification,
and trigger efficiencies separately and then combining them assuming the overall efficiency equal to
the product of the individual components.

The direct measurement is performed in a CR requiring exactly two jets and $100 < \ptmiss < 150\GeV$, excluding the
dilepton invariant mass range $70 < \mll < 110\GeV$ to reduce contributions from \dyjets backgrounds. Here, \Rsfof is
computed using the observed yield of SF and DF events, $\Rsfof = \text{N}_{\text{SF}}/\text{N}_{\text{DF}}$.
Data and simulation agree within 2\% in this region.
In simulation we find that \Rsfof differs by 1\% when computed in the SR instead of the CR.
We check the dependence of \Rsfof on the main kinematic variables used for the analysis in both data
and simulation. Since the statistical power in data is limited, a systematic uncertainty of 4\% is assigned
based on the variations observed in simulation. The measured value of \Rsfof is $1.107 \pm 0.046$.

For the factorized approach, the ratio of muon to electron reconstruction and identification efficiencies, \rmue,
is measured in a \dyjets-enriched CR requiring at least two jets, $\ptmiss < 50\GeV$, and $60 < \mll < 120\GeV$.
This results in a large sample of \ElEl and \MuMu events with similar kinematic distributions to those of the SR.
Assuming the factorization of lepton efficiencies in an event, the efficiency ratio is measured as
$\rmue = \sqrt{\smash[b]{N_{\Pgmp\Pgmm}/N_{{\Pep}{\Pem}}}}$.
This ratio depends on the lepton \pt due to trigger and reconstruction efficiency differences, especially at low lepton \pt.
A parameterization as a function of the \pt of the less energetic lepton is used, and the functional form below
is found to empirically describe the data:

\begin{equation*}
    \rmue  = \rmuec +  \frac{\alpha}{\pt}.
\label{eq:rMuEFormular}
\end{equation*}

Here \rmuec and $\alpha$ are constants that are determined from a fit to data and checked using simulation.
These fit parameters are determined to be $\rmuec =  1.140 \pm 0.005$ and $\alpha = 5.20 \pm 0.16\GeV$.
In addition to the fit uncertainty, a 10\% systematic uncertainty is assigned to account for remaining
variations observed when studying the dependence of \rmue on the \pt of the more energetic lepton, \ptmiss, and the jet multiplicity.

The trigger efficiencies for the three flavor combinations are used to define the factor
$\RT = \sqrt{\smash[b]{\epsilon^{\text{T}}_{\MuMu}\epsilon^{\text{T}}_{\ElEl}}/\epsilon^{\text{T}}_{\Pe^{\pm}\Pgm^{\mp}}}$,
which takes into account the difference between SF and DF channels. The efficiencies
are estimated from a control sample of events collected with a set of nonoverlapping triggers and range between 90--96\%, yielding a final
value of $\RT = 1.052 \pm 0.043$.

The final correction is $\Rsfof = (1/2)(r_{\Pgm\mathrm{/}\Pe}+r_{\Pgm{/}\Pe}^{-1}) \RT$.
The correction relies on the assumption that the number of produced DF events is twice the
number of produced events in each SF sample. Thus, the number of observed DF events needs to be multiplied by
$0.5 \rmue \RT$ and $0.5 \rmue^{-1} \RT$ to predict the number of dimuon and dielectron from FS processes, respectively.
Summing \rmue with its inverse leads to a large reduction in the associated uncertainty. Since \rmue depends
on the lepton kinematic variables, this correction is performed on an event-by-event basis. A separate correction is determined for each SR
and combined with the correction from the direct measurement using the weighted average.

In the method described above, the statistical uncertainty in the predicted number of events is driven
by the statistical uncertainty in the number of data events in the DF CR.
Since \Rsfof is approximately one, the CR yield will be comparable to that of the FS background in the corresponding SR.
In the on-\PZ SRs, the FS background is significantly reduced by the requirement that \mll\ lies within 5\GeV\ of the \PZ\ boson mass.
The expected FS background yields in the SRs are often of the order of a few events or less.
We therefore modify the prediction method to obtain greater statistical power by relaxing the requirement on \mll for DF events,
thereby increasing the number of events in the DF CR.
An additional multiplicative factor, $\kappa$, is calculated and multiplied together with \Rsfof in order to translate this
into a prediction for the SF SR.
The factor $\kappa$ is defined as the number of DF events with
$\abs{m_{\PZ} - \mll} < 5 \GeV$ divided by the number of DF events with $\mll > 20\GeV$.
It is determined from an DF control sample in simulation and validated in the DF data CRs.
A value of $\kappa = 0.065$ is measured from simulation. A systematic uncertainty of 30\% is assigned
by computing $\kappa$ in simulation for both the various on-\PZ SRs
and bins of \ptmiss. The largest observed difference from the nominal $\kappa$ value is taken as the systematic uncertainty.
The value of $\kappa$ derived in data agrees with the result derived from simulation within the assigned uncertainty,
and the statistical uncertainty in the derivation of $\kappa$ is negligible in comparison with the systematic uncertainty.

\subsection{Drell--Yan+jets backgrounds}
\label{sub:dybkg}

The \ptmiss from the \dyjets background is estimated from a sample of photon events in data using
the \ptmiss ``templates'' method~\cite{OSpaperCMS2011,OSpaperCMS7TeV,CMS:edge,CMS:Zedge2015}.
The main premise of this method is
that \ptmiss in \dyjets events originates from the limited detector resolution when measuring the
objects making up the hadronic system that recoils against the \PZ boson.
The shape of the \ptmiss distribution can be estimated from a control sample of \gjets events where the
jet system recoils against a photon instead of a \PZ boson.
In addition to capturing the same resolution effects present in \dyjets events,
the \gjets sample contains more events because of the branching fraction of $\PZ \to \ell^+\ell^-$,
and it does not have any contamination from signal events in the models considered.
For SRs requiring at least one b-tagged jet, some of the observed \ptmiss can originate from neutrinos in semileptonic b-quark decays.
To account for this effect, the \ptmiss templates are extracted from a control sample of \gjets events with the
same b-tagging requirements as in each SR.

The \gjets events in data are selected with a suite of single-photon triggers with \pt thresholds varying from 22 to 165\GeV.
The triggers with thresholds below 165\GeV are prescaled such that only a fraction of accepted events are recorded,
and the events are weighted by the trigger prescales to match the integrated luminosity collected with the signal dilepton triggers.
In order to account for kinematic differences between the hadronic systems in the \gjets and the \dyjets samples,
the \gjets sample is reweighted such that the photon \pt distribution matches the \PZ \pt distribution in the \dyjets sample.
A separate photon CR is defined for each of the on-\PZ SRs in Table~\ref{tab:selections_signalRegions},
where the same kinematic requirements are applied to the \gjets samples as in each SR.
The reweighting in boson \pt is performed for each SR.
Contributions to the photon sample from other SM processes with genuine \ptmiss from prompt neutrinos are subtracted as described below.
The resulting \ptmiss distribution in each SR is then normalized to the observed dilepton data yield
in the range $50 < \ptmiss < 100\GeV$, where \dyjets is the dominant background, after subtracting other background components.

The variable \mttwo\ used in the SR requires two visible objects as input
and thus cannot be calculated in the same way in the photon sample.
Instead, we emulate this requirement in \gjets\ by simulating the decay of the photon to two leptons.
The decay is performed assuming the mother particle has the
mass of a \PZ boson and the momentum of the photon reconstructed from data.
We first consider a system of reference in which the mother particle is at rest.
The decay to the leptons is performed in this system accounting for the
angular dependence of spin correlations in the matrix element.
Then a Lorentz transformation is applied to the emulated dilepton system
in order to match the original momentum of the photon.
The analysis requirements on \pt and $\eta$ for leptons are applied to the simulated decay products.
The variable \mttwol\ is constructed using these leptons, showing good agreement with the distribution of \mttwol\ in genuine \dyjets events,
and a selection is applied to this variable matching each SR requirement.

After selecting events with a high-\pt photon and large \ptmiss,
events from EW processes with genuine \ptmiss, e.g. $\PW\cPgg$ where the \PW\ boson decays to $\ell\nu$,
can be present in the tail of the \ptmiss distribution.
To reduce the contamination from these EW processes, events in the photon sample are removed if they contain a lepton
fulfilling the veto selections for the on-\PZ regions described in Section~\ref{subsec:signalregions}.
We then subtract the residual EW contamination using simulation.
The relative size of the subtraction grows with increasing \ptmiss to be as large as around 50\% of the prediction
or 1 predicted event in the highest \ptmiss bins.

To validate the modeling of the subtracted EW processes, we define a data CR
by selecting events with exactly one muon and one photon,
requiring $\ptmiss > 50\GeV$ and the transverse mass \mt\ of the muon and \ptmiss to be greater than 30\GeV.
The muon must satisfy $\pt > 25\GeV$, and the events are selected using a trigger that requires
at least one isolated muon with $\pt > 24\GeV$.  This region consists of about 50\% $\PW\cPgg$ events
with the remainder coming primarily from $\ttbar\cPgg$ events.  Agreement is observed between data and the prediction from simulation.
Based on the level of agreement between data and simulation in the kinematic distributions of photon \pt\ and \ptmiss,
we assign a systematic uncertainty of 30\% in the subtraction of these EW processes.

The systematic uncertainty in the prediction takes into account the statistical uncertainty in the \gjets sample
in each bin of \ptmiss, which is the dominant uncertainty in the highest \ptmiss bins.
The statistical uncertainty in the normalization region of $50 < \ptmiss < 100\GeV$ is included and ranges from 7--30\%.
A closure test of the method is performed in simulation, using \gjets to predict the yield of \dyjets in each analysis bin.
An uncertainty is assigned from
the results of this test as the larger of the difference between the \gjets prediction and the \dyjets yield for each \ptmiss region
or the simulation statistical uncertainty. The values vary between 10 and 80\% depending on the \ptmiss region,
with the larger values coming from regions with low statistics in simulation.

The template method is also used to provide a prediction for the background from \dyjets
in the edge SRs, where this background is significantly smaller due to the \mll\ requirements.
We define the ratio \rinout in a \dyjets-dominated sample as the number of SF events in a given bin of \mll
divided by the SF yield within $86 < \mll < 96\GeV$.
The ratio is measured in a \dyjets-dominated
CR requiring at least two jets, $\ptmiss < 50\GeV$, and $\mttwol >80\GeV$.
Different-flavor yields in both the numerator and denominator are subtracted from the respective SF yields in order to correct
for small FS contributions in the region where \rinout is measured.
The value of \rinout ranges from 0.001 to 0.16 for the different bins in \mll.
The \dyjets background contribution to each \mll bin is computed by multiplying the on-\PZ prediction by \rinout.
The dependence of \rinout on \ptmiss and the jet multiplicity are studied in the data CR.
Based on the statistical precision of this check, and the observed variations as a function of these variables,
we assign an uncertainty of 50 (100)\% to \rinout in the \mll bins below (above) 150\GeV.

\subsection{Backgrounds with \texorpdfstring{\PZ}{Z} bosons plus genuine \texorpdfstring{\ptmiss}{missing transverse momentum}}
\label{sub:zmetbkg}
The \ptmiss\ template method only predicts instrumental \ptmiss\ from jet mismeasurement
and thus does not include the genuine \ptmiss\ from prompt neutrinos expected in processes like $\PW(\ell\nu)\PZ(\ell\ell)$,
$\PZ(\ell\ell)\PZ(\nu\nu)$, or lower cross section processes such as \ttz.
These processes can be a substantial fraction of the background at high \ptmiss and are estimated using simulation.

The prediction from simulation is validated by comparing to data in CRs requiring three or four leptons.
A region enriched in \PW\PZ events is selected by requiring exactly three leptons, at least two jets, no b-tagged jets, $\ptmiss > 60\GeV$,
and an OCSF lepton pair with $86 < \mll < 96\GeV$.
Another three-lepton CR is defined targeting \ttz by requiring at least two jets, at least two b-tagged jets, $\ptmiss > 30\GeV$,
and an OCSF lepton pair as in the \PW\PZ region.
A four-lepton CR targeting $\PZ\PZ$ is constructed by requiring four leptons with two OCSF pairs satisfying $\mll > 20\GeV$,
to remove low-mass resonances, and at least two jets.

After subtracting the other processes using simulation in each region, simulation-to-data scale factors of
$0.98 \pm 0.11$, $1.58 \pm 0.49$, and $1.31 \pm 0.29$ are observed for \PW\PZ, $\PZ\PZ$, and \ttz backgrounds respectively.
We use the scale factor values to correct the prediction from simulation for each process.
Based on the statistical uncertainty in these CRs, and the agreement between data and simulation in distributions of
kinematic variables such as \ptmiss\ and the number of jets, we assign systematic uncertainties of 30\%
for the \PW\PZ and \ttz background predictions and 50\% for the $\PZ\PZ$ prediction.
As all other effects are subdominant, we do not assign further uncertainties
to these backgrounds.

\section{Kinematic fit}
\label{sec:kinfit}

A simultaneous extended unbinned maximum likelihood fit is performed in the \mll\ distributions of \EE, \MM, and \EM events
to search for a kinematic edge.
The fit is performed after the kinematic selection labeled ``Edge fit'' in Table~\ref{tab:selections_signalRegions}.
The likelihood model contains three components:
an FS background component, a \dyjets background component, and a signal component.
The \znu background is contained within the \dyjets component in this method, as both have the same \mll\ shape.

The FS background component is described using a CB function $\mathcal{P}_{\mathrm{CB}}(\mll)$:
\begin{equation}
\begin{aligned}
\mathcal{P}_{\mathrm{CB}}(\mll) = \begin{cases}
\exp\left[-\frac{(\mll-\mu_{\mathrm{CB}})^2}{2\Gamma_{\mathrm{CB}}^2}\right] &\text{ if } \frac{\mll-\mu_{\mathrm{CB}}}{\Gamma_{\mathrm{CB}}}<\alpha \\
A (B+\frac{\mll-\mu_{\mathrm{CB}}}{\Gamma_{\mathrm{CB}}})^{-n} &\text{ if } \frac{\mll-\mu_{\mathrm{CB}}}{\Gamma_{\mathrm{CB}}}>\alpha \\
\end{cases}
\end{aligned}
\end{equation}
where
\begin{equation}
A = \left(\frac{n}{|\alpha|}\right)^{n} \exp\left(-\frac{|\alpha|^2}{2}\right) \quad \text{and}\quad B = \frac{n}{|\alpha|}-|\alpha|.
\end{equation}
The FS background model has five free parameters: the overall normalization, the mean $\mu_{\mathrm{CB}}$ and width $\Gamma_{\mathrm{CB}}$ of the Gaussian part, the transition point $\alpha$ between the Gaussian part
and the power law tail, and the power law parameter $n$.

The \dyjets background component is modeled with the sum of an exponential function, which describes the low-mass rise, and a Breit--Wigner
function with a mean and width set to the nominal \PZ boson values~\cite{Olive:2016xmw}, which accounts for the \PZ boson lineshape.
To account for the experimental resolution, the Breit--Wigner function is convolved with a double-sided CB function
\begin{equation}
\begin{aligned}
\mathcal{P}_{\mathrm{DSCB}}(\mll) = \begin{cases} A_{1} (B_{1}-\frac{\mll-\mu_{\mathrm{DSCB}}}{\Gamma_{\mathrm{DSCB}}})^{-n_{1}} &\text{ if } \frac{\mll-\mu_{\mathrm{DSCB}}}{\Gamma_{\mathrm{DSCB}}}<-\alpha_{1} \\
\exp\left[-\frac{(\mll-\mu_{\mathrm{DSCB}})^2}{2\Gamma_{\mathrm{DSCB}}^2}\right] &\text{ if } -\alpha_{1}<\frac{\mll-\mu_{\mathrm{DSCB}}}{\Gamma_{\mathrm{DSCB}}}<\alpha_{2} \\
A_{2} (B_{2}+\frac{\mll-\mu_{\mathrm{DSCB}}}{\Gamma_{\mathrm{DSCB}}})^{-n_{2}} &\text{ if } \frac{\mll-\mu_{\mathrm{DSCB}}}{\Gamma_{\mathrm{DSCB}}}>\alpha_{2} \\
\end{cases}
\end{aligned}
\end{equation}
where $\mu_{\mathrm{DSCB}}$ and $\Gamma_{\mathrm{DSCB}}$ are the mean and width, respectively, of the CB function,
and $\alpha_{1}$ and $\alpha_{2}$ are the transition points.
The full model for the on-\PZ \dyjets background line shape is thus
\begin{equation}
\mathcal{P}_{\mathrm{DY},\text{ on-Z }} (\mll) = \int \mathcal{P}_{\mathrm{DSCB}}(\mll)\mathcal{P}_{\mathrm{BW}}(\mll-m') \rd{}m',
\end{equation}
where $\mathcal{P}_{\mathrm{BW}}$ is the Breit--Wigner function. The complete \dyjets background model has nine free parameters.

The signal component is described by a triangular shape, convolved with a Gaussian distribution to account for the experimental resolution:
\begin{equation}
 {\mathcal{P}}_{\mathrm{S}}(\mll) \propto \frac{1}{\sqrt{2\pi}\Gamma_{\ell\ell}} \int_{0}^{\mll^{\text{edge}}} y \exp\left[ -\frac{(\mll-y)^2}{2\Gamma_{\ell\ell}^{2}}\right]\, \rd{}y.
\end{equation}
The signal model has two free parameters: the fitted signal yield and the position of the edge, $\mll^{\text{edge}}$.

As the first step, a fit is performed separately for \EE and \MM events in a \dyjets-enriched CR requiring at least two jets and
$\ptmiss < 50\GeV$, to determine the shape of backgrounds containing a \PZ boson.
The parameters of the \dyjets background shape are then fixed and only the normalizations of these backgrounds are free parameters in the subsequent fit.
The final fit is performed simultaneously to the dilepton invariant mass distributions in the \EE, \MM, and \EM~samples.
The model for the FS background is the same for the SF and DF events.
The \Rsfof factor is treated as a nuisance parameter, parameterized by a Gaussian distribution
with a mean value and standard deviation given by the value of
\Rsfof and its uncertainties (see Section~\ref{sub:fsbkg}). The final fit has ten free parameters: a normalization for each
of the three fit components, four parameters for the shape of the FS background, \Rsfof, the relative fraction of
dielectron and dimuon events in the FS prediction, and the position of the signal edge.

\section{Results}
\label{sec:results}

The observed number of events in the SRs are compared with the background estimates
for the on-\PZ strong- and EW-production and the edge searches. The
covariance and correlation matrices of the background predictions in the different SRs are also provided in Appendix~\ref{sec:appA}
to facilitate reinterpretation of these results.
For the edge search, the fit is performed to search for a kinematic edge in the \mll spectrum.

\subsection{Results of the search in the \texorpdfstring{on-\PZ}{on-Z} signal regions}
\label{sub:onZResults}

\begin{figure}[htbp]
\centering
\includegraphics[width=0.42\linewidth]{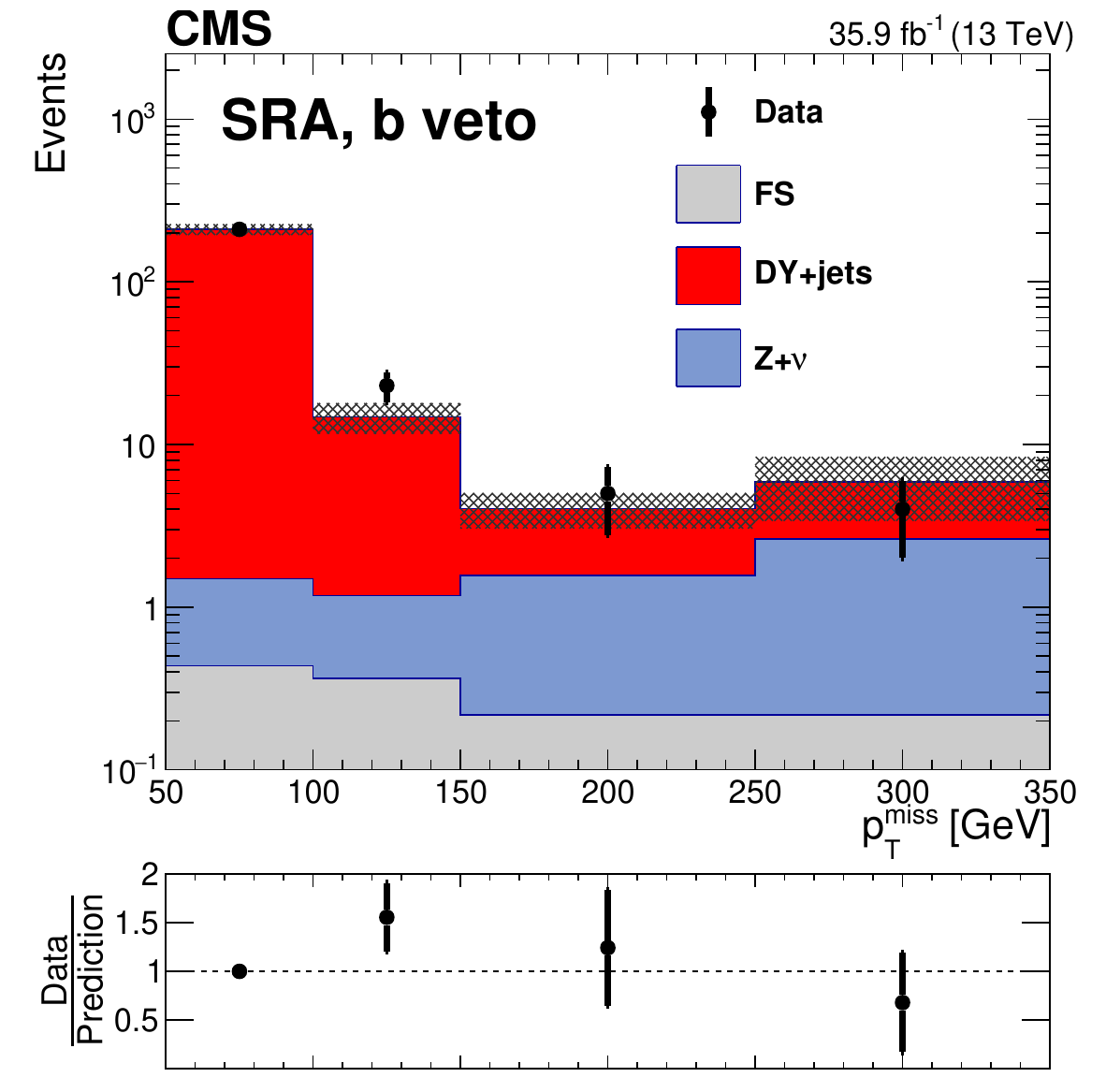}
\includegraphics[width=0.42\linewidth]{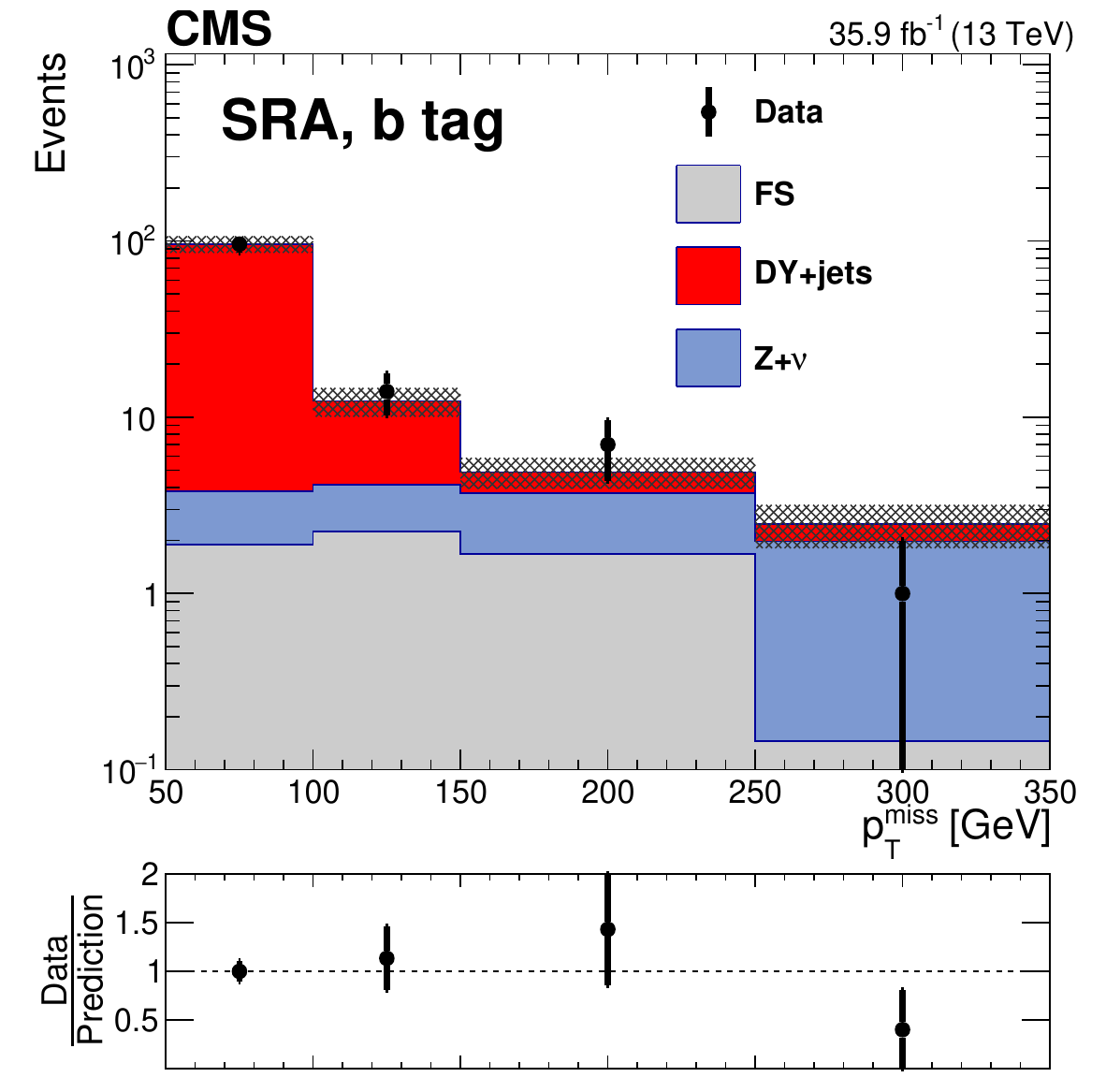}
\includegraphics[width=0.42\linewidth]{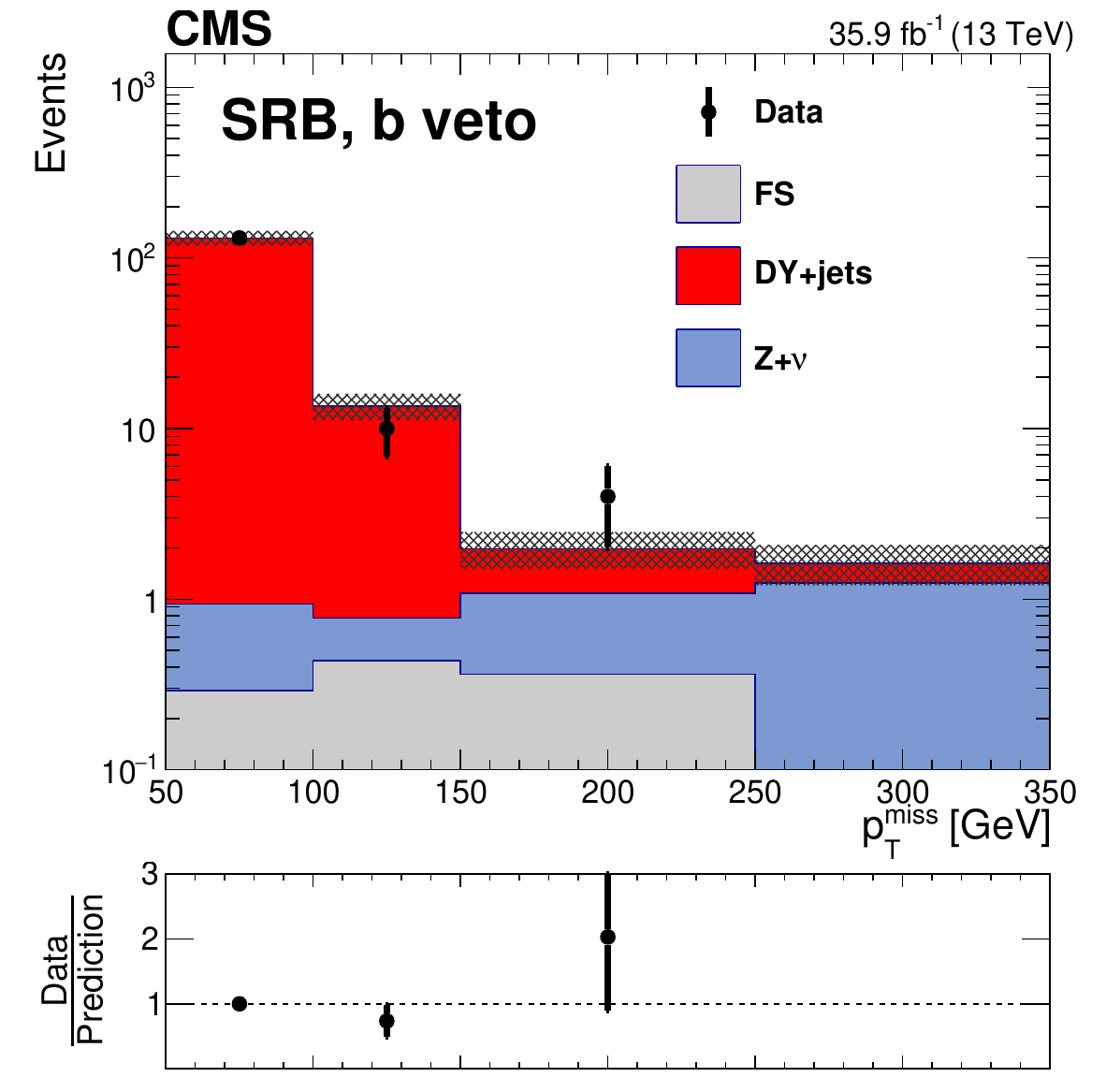}
\includegraphics[width=0.42\linewidth]{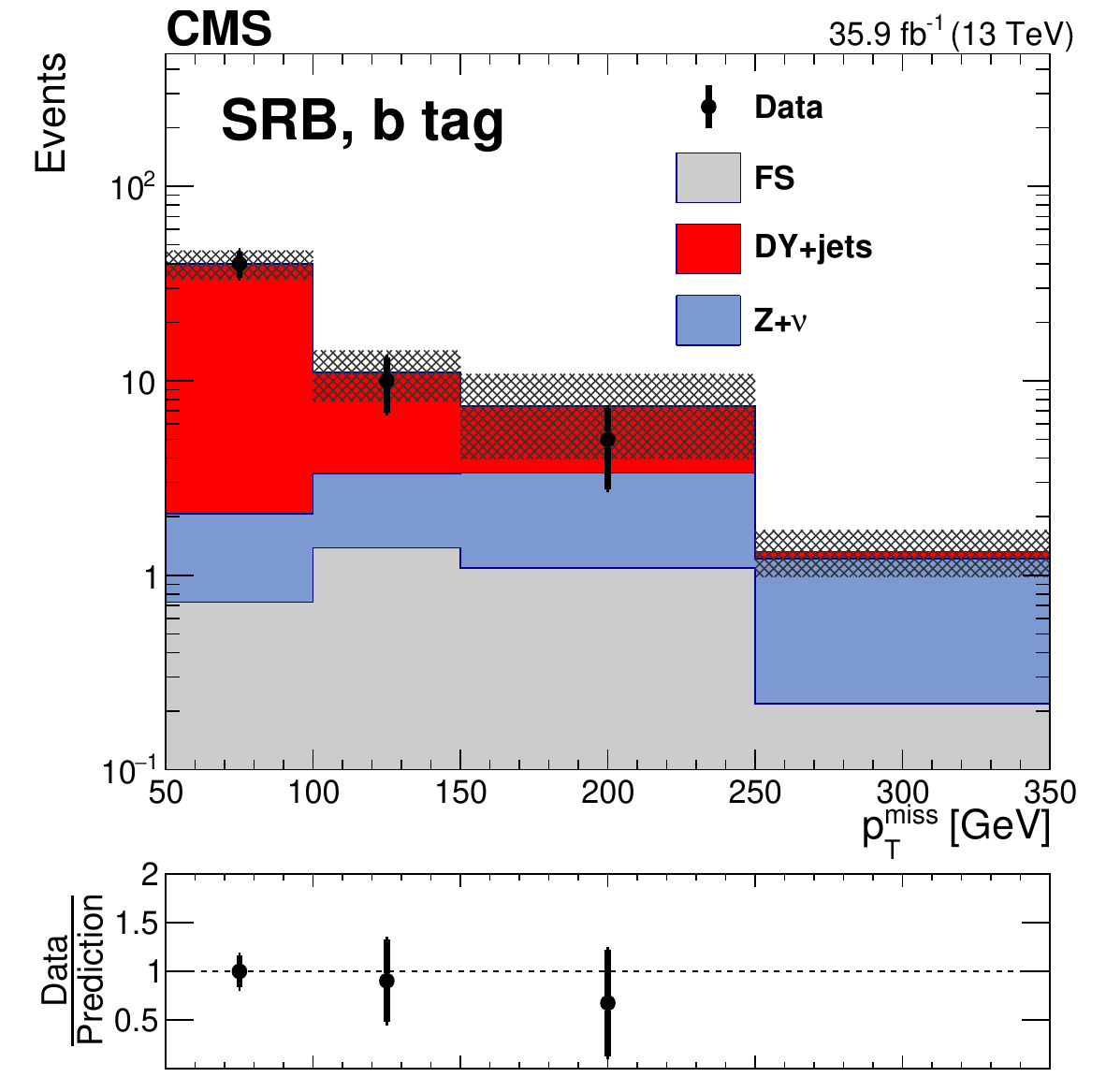}
\includegraphics[width=0.42\linewidth]{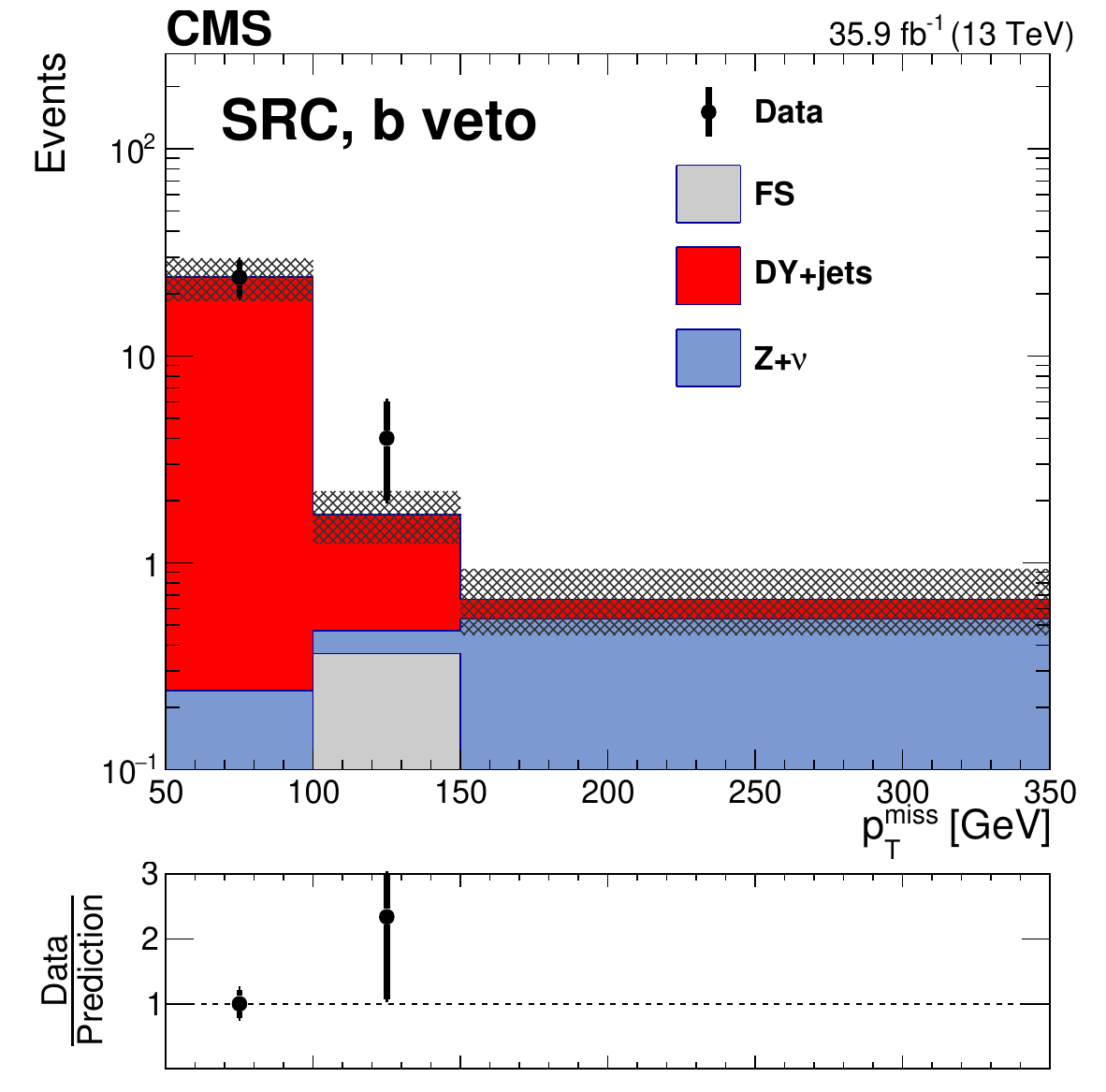}
\includegraphics[width=0.42\linewidth]{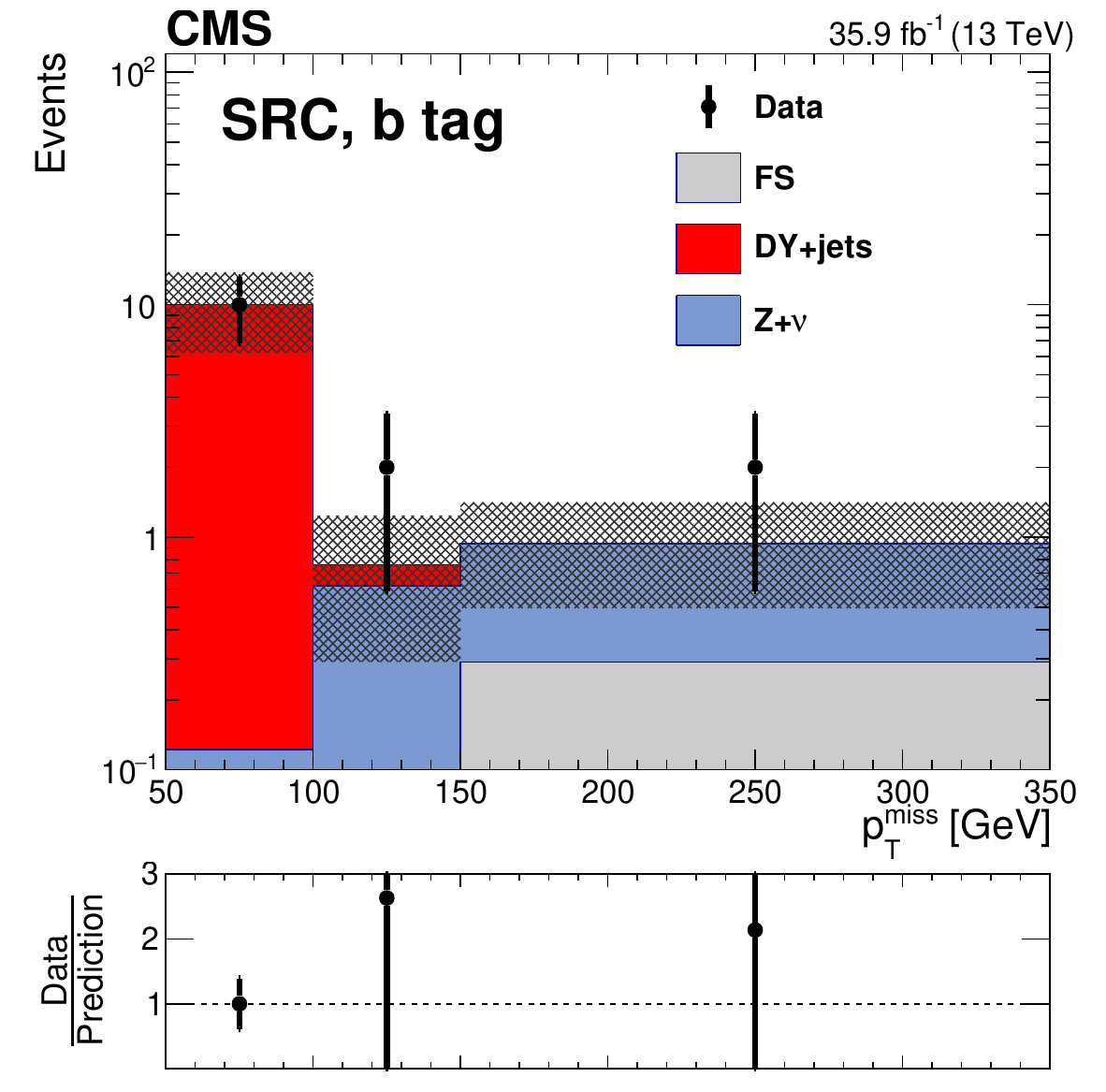}
\caption{\label{fig:results_SR_str}
  The \ptmiss\ distribution is shown for data compared to the background prediction in the on-\PZ strong-production SRs
  with no b-tagged jets (left) and at least 1 b-tagged jet (right).
  The rows show SRA (upper), SRB (middle), and SRC (lower).
  The lower panel of each plot shows the ratio of observed data to the predicted value in each bin.
  The hashed band in the upper panels shows the total uncertainty in the background prediction,
  including statistical and systematic components.
The \ptmiss template prediction for each SR is normalized to the first bin of each distribution,
and therefore the prediction agrees with the data by construction.
}
\end{figure}

The results for the SRs of the on-\PZ strong-production search are presented in Table~\ref{tab:results_SR_str}.
The corresponding \ptmiss\ distributions are shown in Fig.~\ref{fig:results_SR_str}.
No significant deviations are observed with respect to SM expectations.

\begin{table}[htbp]
\centering
\topcaption{\label{tab:results_SR_str}
  Predicted and observed event yields are shown for the on-\PZ strong-production SRs, for each \ptmiss\ bin defined
  in Table~\ref{tab:selections_signalRegions}.
The uncertainties shown include both statistical and systematic components.
}
\begin{tabular} {l  l  c c c }
\hline
SRA, b veto & \ptmiss [\GeVns{}]& 100--150              & 150--250                                      & $>$250 \\ \hline
            & \dyjets        & 13.6$\pm$3.1\x         & 2.5$\pm$0.9                                   & 3.3$\pm$2.4 \\
            & FS            & $0.4^{+0.3}_{-0.2}$\x  & $0.2^{+0.2}_{-0.1}$\x                           & $0.2^{+0.2}_{-0.1}$\x  \\
            & \znu          & 0.8$\pm$0.3          & 1.4$\pm$0.4                                   & 2.4$\pm$0.8 \\
            & Total background           & 14.8$\pm$3.2\x & 4.0$\pm$1.0                           & 5.9$\pm$2.5 \\
            & Data          & 23                   & 5                                             & 4 \\ \hline

SRA, b tag  & \ptmiss [\GeVns{}]& 100--150              & 150--250                                      & $>$250 \\ \hline
            & \dyjets        & 8.2$\pm$2.1          & 1.2$\pm$0.5                                   & 0.5$\pm$0.3 \\
            & FS            & 2.3$\pm$0.8  & $1.7^{+0.7}_{-0.6}$\x                           & $0.1^{+0.2}_{-0.1}$\x  \\
            & \znu          & 1.9$\pm$0.4          & 2.0$\pm$0.5                                   & 1.8$\pm$0.6 \\
            & Total background           & 12.4$\pm$2.3\x & 4.9$\pm$1.0                           & 2.5$\pm$0.7 \\
            & Data          & 14                   & 7                                             & 1 \\ \hline

SRB, b veto & \ptmiss [\GeVns{}]& 100--150              & 150--250                                      & $>$250 \\ \hline
            & \dyjets        & 12.8$\pm$2.3\x         & 0.9$\pm$0.3                                   & 0.4$\pm$0.2 \\
            & FS            & $0.4^{+0.3}_{-0.2}$\x  & $0.4^{+0.3}_{-0.2}$\x                           & $0.1^{+0.2}_{-0.1}$\x  \\
            & \znu          & 0.3$\pm$0.1          & 0.7$\pm$0.2                                   & 1.2$\pm$0.4 \\
            & Total background           & 13.6$\pm$2.4\x & 2.0$\pm$0.5                           & 1.6$\pm$0.4 \\
            & Data          & 10                   & 4                                             & 0 \\ \hline

SRB, b tag  & \ptmiss [\GeVns{}]& 100--150              & 150--250                                      & $>$250 \\ \hline
            & \dyjets        & 7.7$\pm$3.2          & 4.0$\pm$3.4                                   & 0.1$\pm$0.1 \\
            & FS            & $1.4^{+0.6}_{-0.5}$\x  & $1.1^{+0.5}_{-0.4}$\x                           & $0.2^{+0.2}_{-0.1}$\x  \\
            & \znu          & 2.0$\pm$0.5          & 2.3$\pm$0.6                                   & 1.0$\pm$0.3 \\
            & Total background           & 11.1$\pm$3.3\x & $7.4^{+3.5}_{-3.4}$\x                           & $1.3^{+0.4}_{-0.3}$\x \\
            & Data          & 10                   & 5                                             & 0 \\ \hline

SRC, b veto & \ptmiss [\GeVns{}]& 100--150              &  $>$150 & \\ \hline
            & \dyjets        & 1.2$\pm$0.4          & { 0.1$\pm$0.1 }& \\
            & FS            & $0.4^{+0.3}_{-0.2}$\x  &{ $0.1^{+0.2}_{-0.1}$\x  }& \\
            & \znu          & 0.1$\pm$0.1          & { 0.5$\pm$0.2 }& \\
            & Total background           & 1.7$\pm$0.5  & { $0.7^{+0.3}_{-0.2}$\x }& \\
            & Data          & 4                    & { 0 }& \\ \hline

SRC, b tag  & \ptmiss [\GeVns{}]& 100--150              & { $>$150 }& \\ \hline
            & \dyjets        & 0.1$\pm$0.4          & { 0.0$\pm$0.3 }& \\
            & FS            & $0.0^{+0.1}_{-0.0}$\x  & { 0.3$\pm$0.2  }& \\
            & \znu          & 0.6$\pm$0.2          & { 0.6$\pm$0.2 }& \\
            & Total background           & 0.8$\pm$0.5  & { $0.9^{+0.5}_{-0.4}$\x }& \\
            & Data          & 2                    & { 2 }& \\ \hline

\end{tabular}
\end{table}

\begin{figure}[tbh]
\centering
\begin{tabular}{cc}
\includegraphics[width=0.45\linewidth]{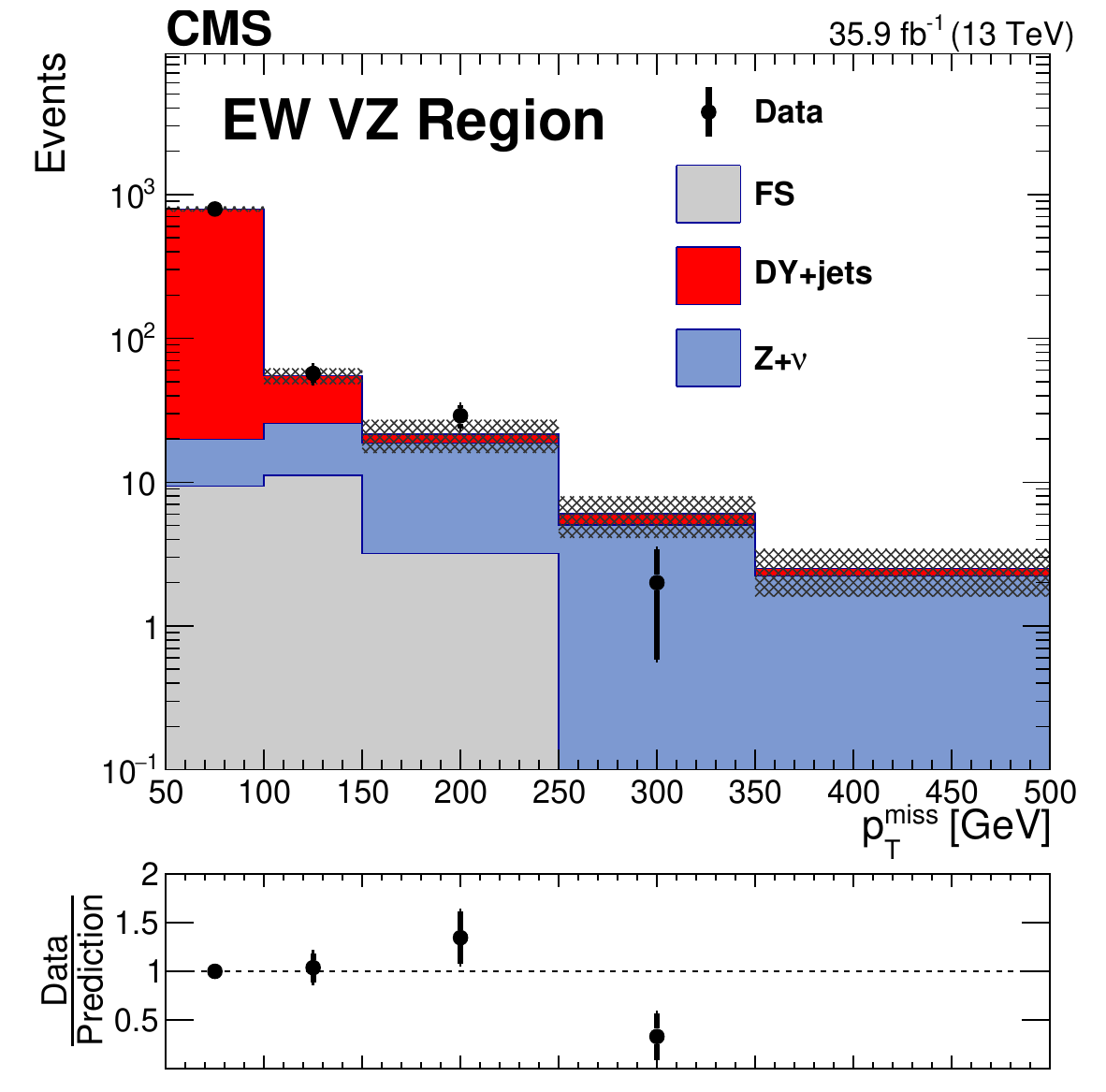} &
\includegraphics[width=0.45\linewidth]{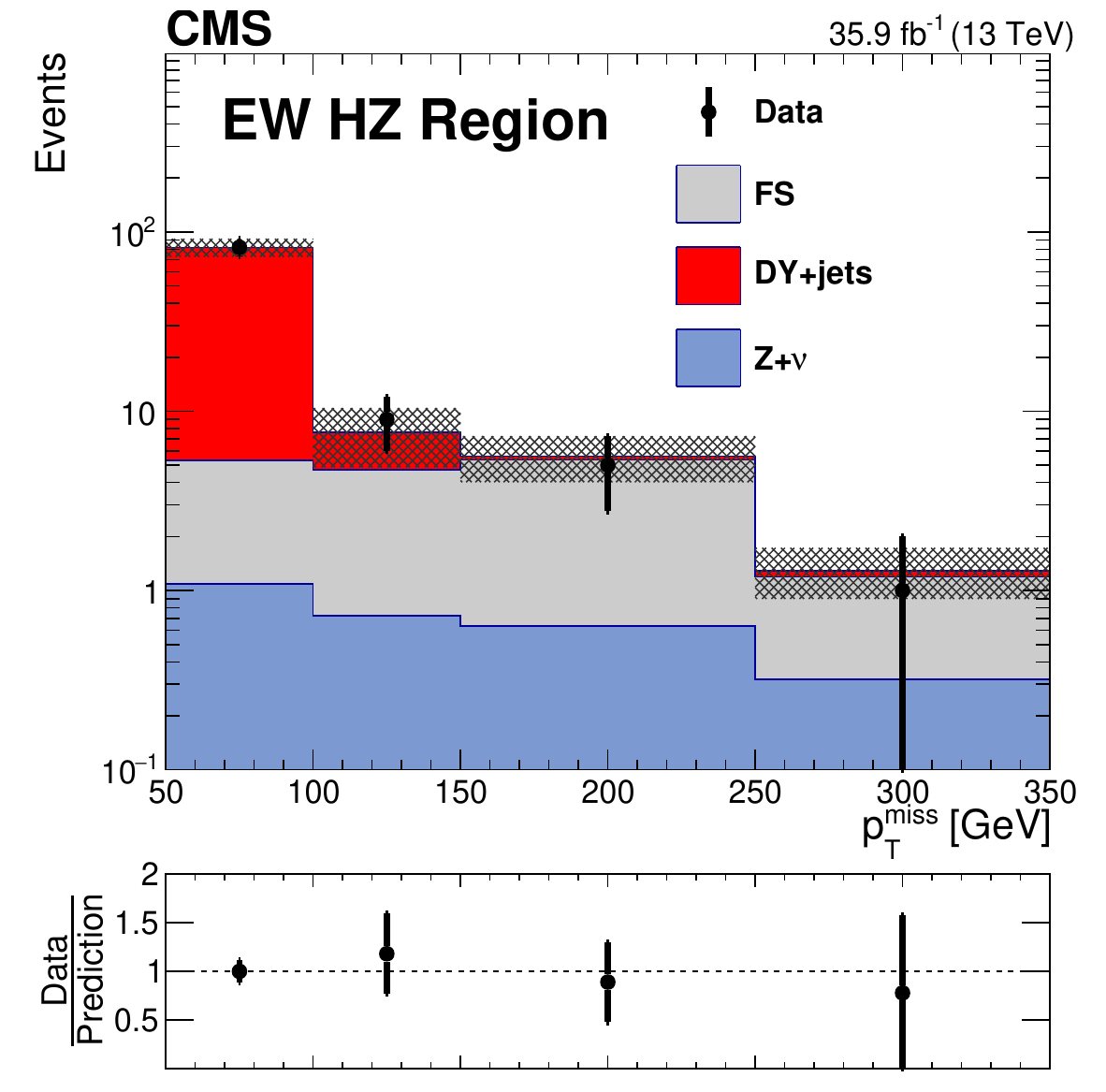} \\
\end{tabular}
\caption{
  The \ptmiss\ distribution is shown for data compared to the background prediction in the on-\PZ \vz (left)
  and $\PH\PZ$ (right) electroweak-production SRs.
  The lower panel of each figure shows the ratio of observed data to the predicted value in each bin.
  The hashed band in the upper panels shows the total uncertainty in the background prediction,
  including statistical and systematic sources.
  The \ptmiss template prediction for each SR is normalized to the first bin of each distribution,
  and therefore the prediction agrees with the data by construction.
\label{fig:results_SR_ewk}
}
\end{figure}

The results for the EW SRs in the on-\PZ search are shown in Table~\ref{tab:results_SR_ewk}.
The corresponding \ptmiss\ distributions are shown in Fig.~\ref{fig:results_SR_ewk}.
The observed data are also consistent with the background prediction.

\begin{table}[tbh]
\centering
\topcaption{\label{tab:results_SR_ewk}
  Predicted and observed event yields are shown for the EW on-\PZ SRs, for each \ptmiss\ bin defined
  in Table~\ref{tab:selections_signalRegions}.
The uncertainties shown include both statistical and systematic sources.
}
\begin{tabular} {l  l  c c c c }
\hline
\vz     & \ptmiss [\GeVns{}]& 100--150              & 150--250             & 250--350                                      & $>$350 \\ \hline
        & \dyjets        & 29.3$\pm$4.4\x         & 2.9$\pm$2.0          & 1.0$\pm$0.7                                   & 0.3$\pm$0.3 \\
        & FS            & 11.1$\pm$3.6\x & 3.2$\pm$1.1  & $0.1^{+0.2}_{-0.1}$\x                           & $0.1^{+0.2}_{-0.1}$\x  \\
        & \znu          & 14.5$\pm$4.0\x         & 15.5$\pm$5.1\x         & 5.0$\pm$1.8                                   & 2.2$\pm$0.9 \\
        & Total background           & 54.9$\pm$7.0\x & 21.6$\pm$5.6\x & 6.0$\pm$1.9                           & 2.5$\pm$0.9 \\
        & Data          & 57                   & 29                   & 2                                             & 0 \\ \hline

$\PH\PZ$ & \ptmiss [\GeVns{}]& 100--150              & 150--250             & { $>$250 } & \\ \hline
        & \dyjets        & 2.9$\pm$2.4          & 0.3$\pm$0.2          & { 0.1$\pm$0.1 } & \\
        & FS            & 4.0$\pm$1.4  & 4.7$\pm$1.6  & { 0.9$\pm$0.4  } & \\
        & \znu          & 0.7$\pm$0.2          & 0.6$\pm$0.2          & { 0.3$\pm$0.1 } & \\
        & Total background           & 7.6$\pm$2.8  & 5.6$\pm$1.6  & { 1.3$\pm$0.4 } & \\
        & Data          & 9                    & 5                    & { 1 } & \\ \hline

\end{tabular}
\end{table}

\subsection{Results of the edge search}
\label{sub:edgeResults}

\begin{figure}[!h]
\centering
\includegraphics[width=0.7\textwidth]{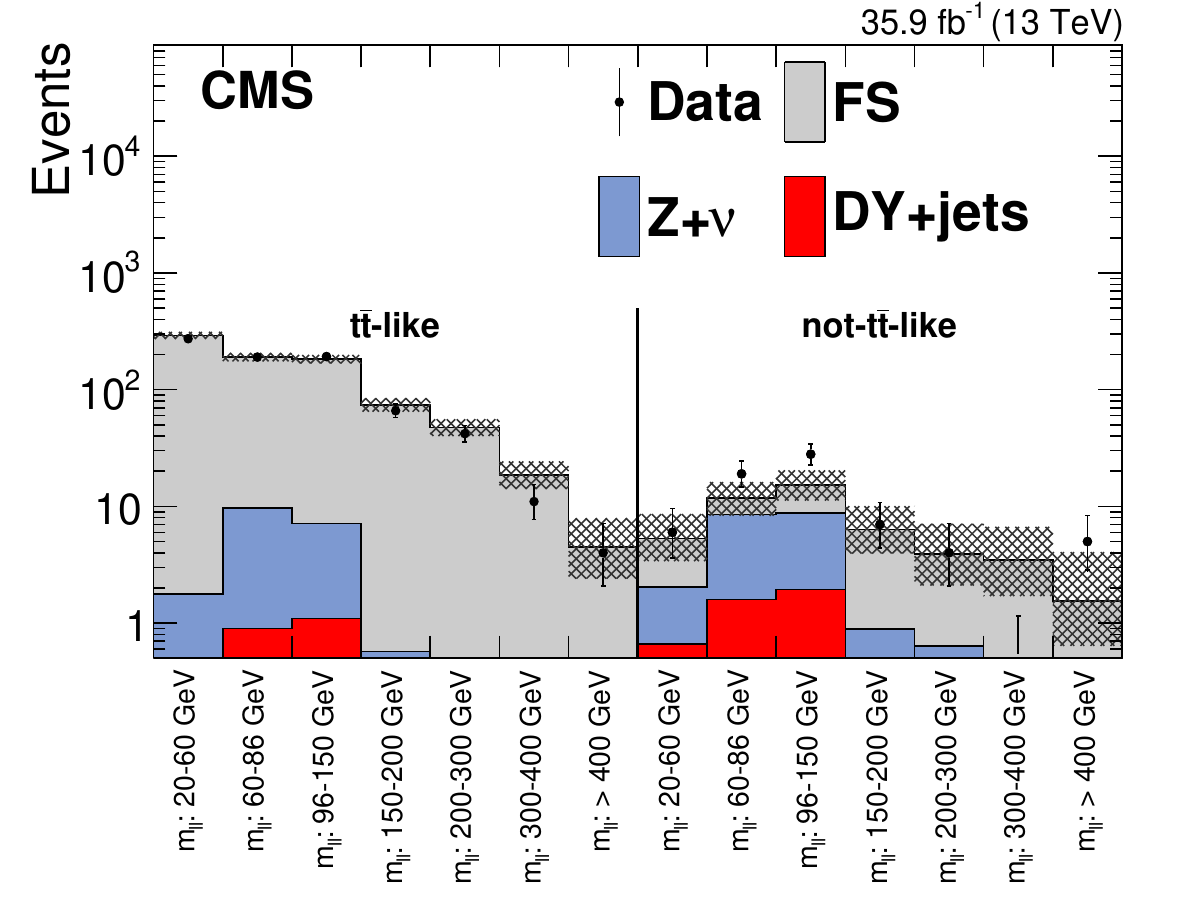}
\caption{Results of the counting experiment of the edge search. For each SR, the number of observed events, shown as black data points,
  is compared to the total background estimate.  The hashed band shows the total uncertainty in the background prediction,
  including statistical and systematic sources.}
\label{fig:cNc_resultOverview}

\end{figure}

The edge search features seven distinct \mll regions, each of which is divided into two bins using the likelihood discriminant,
resulting in fourteen SRs. In
addition, two aggregate regions integrating the SRs below and above the \PZ boson mass have been considered in the not-\ttbar-like case.
Table~\ref{tab:edgeResults} summarizes the SM predictions and the observations in these SRs. A graphical representation of these results
is shown in Fig.~\ref{fig:cNc_resultOverview}, including the relative contributions of the different backgrounds.

\begin{table}[tbh]
\renewcommand{\arraystretch}{1.3}
\setlength{\belowcaptionskip}{6pt}
\centering
\topcaption{Predicted and observed yields in each bin of the edge search counting experiment.
The uncertainties shown include both statistical and systematic sources.
}
\label{tab:edgeResults}
\begin{tabular}{ c  c  c  c  c  c}
\hline
\mll range [\GeVns{}]& FS & \dyjets & \znu  & Total background & Data\\
\hline
\multicolumn{6}{c}{\ttbar-like}  \\
\hline
20--60   &  291$^{+21}_{-20}$    & 0.4$\pm$0.3   & 1.4$\pm$0.5  &  293$^{+21}_{-20}$ & 273 \\
60--86   &  181$^{+16}_{-15}$    & 0.9$\pm$0.7   & 8.8$\pm$3.4  &  190$^{+16}_{-15}$ & 190 \\
\x96--150   &  176$^{+15}_{-14}$    & 1.1$\pm$0.9   & 6.0$\pm$2.4  &  182$^{+16}_{-15}$ & 192 \\
150--200   &  \x73$^{+10}_{-9}$    & 0.1$\pm$0.1   & 0.4$\pm$0.2  &  \x74$^{+10}_{-9}$ & 66 \\
200--300   &  46.9$^{+8.4}_{-7.3}$\y    & $<$0.1    & 0.3$\pm$0.1  &  47.3$^{+8.4}_{-7.3}$\y & 42 \\
300--400   &  18.5$^{+5.7}_{-4.5}$\y    & $<$0.1   & $<$0.1  &  18.6$^{+5.7}_{-4.5}$\y & 11 \\
$>$400   &  \x4.3$^{+3.4}_{-2.1}$\y    & $<$0.1   & $<$0.1  &  \x4.5$^{+3.4}_{-2.1}$\y & 4 \\
\hline
\multicolumn{6}{c}{Not-\ttbar-like}   \\
\hline
20--60   &  3.3$^{+3.2}_{-1.8}$    & 0.7$\pm$0.5   & 1.4$\pm$0.5  &  5.3$^{+3.3}_{-1.9}$ & 6 \\
60--86   &  3.3$^{+3.2}_{-1.8}$    & 1.6$\pm$1.3   & 6.9$\pm$2.7  &  11.8$^{+4.4}_{-3.5}$\x & 19 \\
\x96--150   &  6.6$^{+3.9}_{-2.6}$    & 1.9$\pm$1.5   & 6.8$\pm$2.7  &  15.3$^{+5.0}_{-4.1}$\x & 28 \\
150--200   &  5.5$^{+3.7}_{-2.4}$    & 0.2$\pm$0.3   & 0.7$\pm$0.3  &  6.4$^{+3.7}_{-2.4}$ & 7 \\
200--300   &  3.3$^{+3.2}_{-1.8}$    & 0.2$\pm$0.2   & 0.5$\pm$0.2  &  3.9$^{+3.2}_{-1.8}$ & 4 \\
300--400   &  3.3$^{+3.2}_{-1.8}$    & $<$0.1    & 0.2$\pm$0.1  &  3.5$^{+3.2}_{-1.8}$ & 0 \\
$>$400   &  1.1$^{+2.5}_{-0.9}$    & $<$0.1    & 0.4$\pm$0.2  &  1.6$^{+2.5}_{-0.9}$ & 5 \\
\hline
\multicolumn{6}{c}{Aggregate SRs (not-\ttbar-like)}  \\
\hline
20--86   &  6.5$^{+3.9}_{-2.6}$    & 2.3$\pm$1.5   & 8.3$\pm$3.2  &  17.1$^{+5.3}_{-4.4}$\y & 25 \\
$>$96   &  19.6$^{+5.8}_{-4.6}$\x    & 2.4$\pm$1.6   & 8.5$\pm$3.4  &  30.6$^{+7.0}_{-6.0}$\x & 44 \\
\hline
\end{tabular}
\end{table}

At high mass and in the not-\ttbar-like regions, the uncertainty in the background prediction is driven by
the statistical uncertainty in the number of events in the DF control sample.
There is good agreement between prediction and observation for all SRs. The largest deviation is observed in the not-\ttbar-like region for masses between 96 and 150\GeV, with an excess corresponding
to a local significance of 2.0 standard deviations.

\begin{figure}[!hbtp]
\centering
\includegraphics[width=0.42\textwidth]{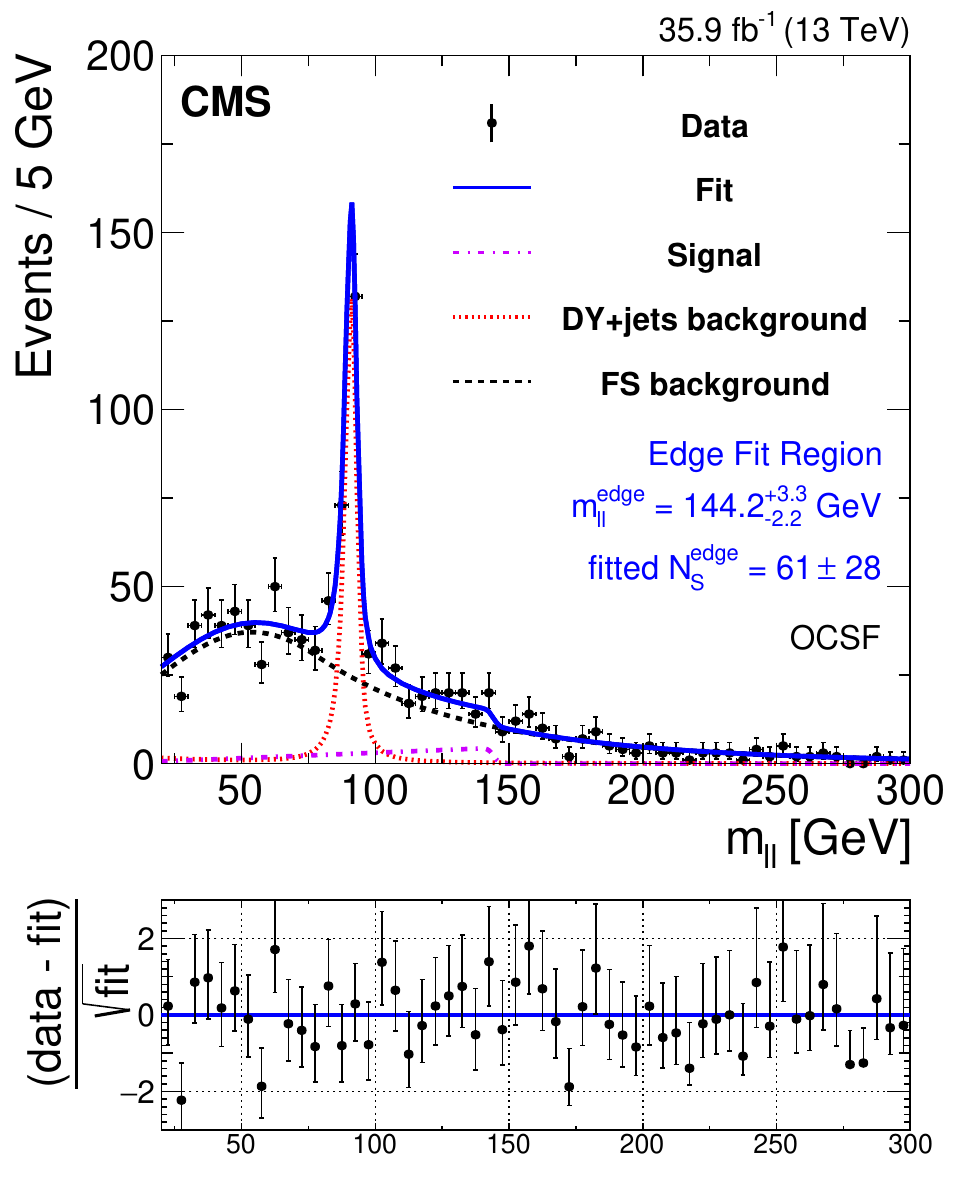}
\includegraphics[width=0.42\textwidth]{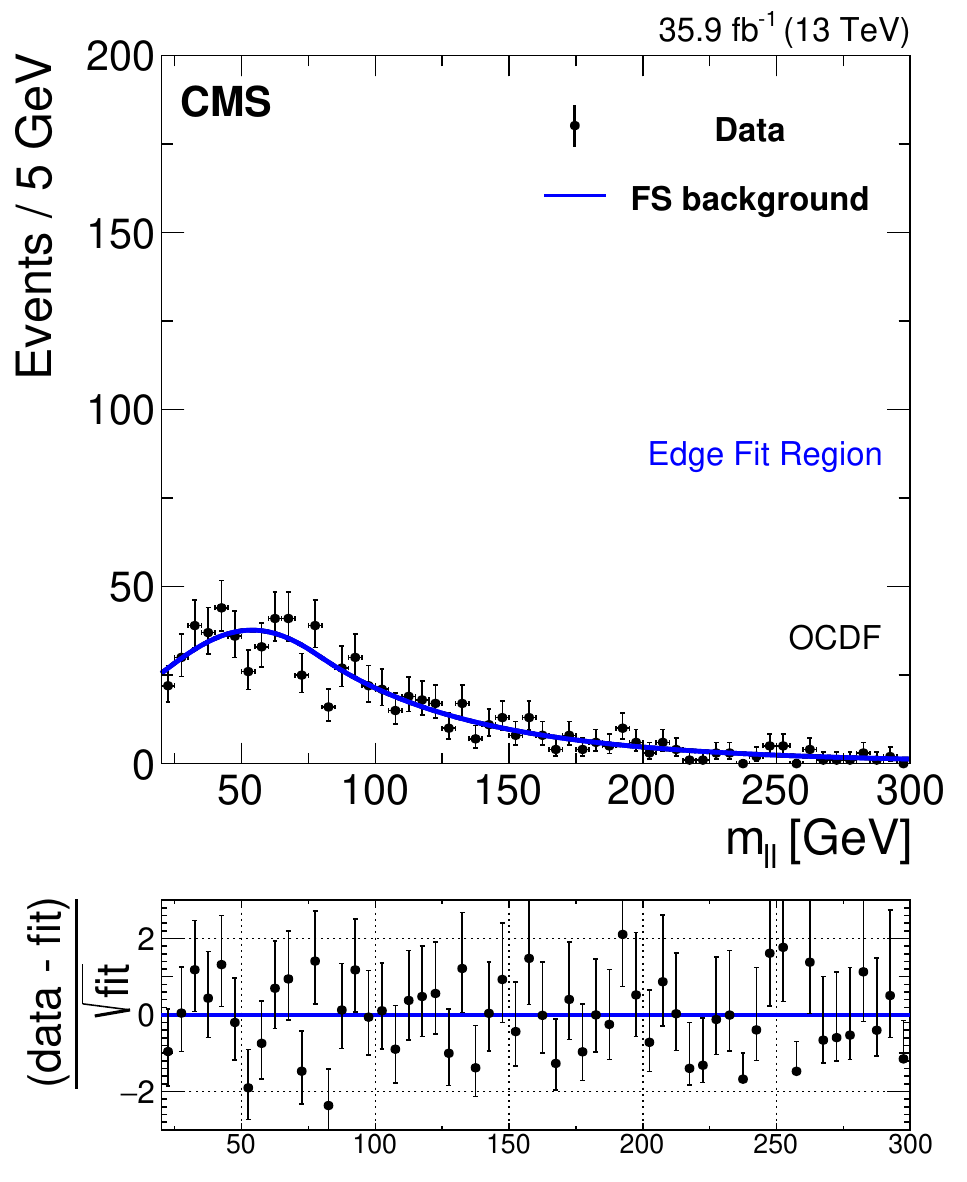}
\caption{
  Fit of the dilepton mass distributions to the signal-plus-background hypothesis
  in the ``Edge fit'' SR from Table~\ref{tab:selections_signalRegions},
    projected on the same-flavor (left) and different-flavor (right) event samples.
    The fit shape is shown as a solid blue line.
    The individual fit components are indicated by dashed and dotted lines.
    The FS background is shown with a black dashed line.
    The \dyjets background is displayed with a red dotted line.
    The extracted signal component is displayed with a purple dash-dotted line.
    The lower panel in each plot shows the difference between the observation and the fit, divided by the square root of the number of fitted events.
}
\label{fig:Fit_data_H1}

\end{figure}

The dilepton mass distributions and the results of the kinematic fit are shown in Fig.~\ref{fig:Fit_data_H1}.
Table~\ref{tab:fitResults} presents a summary of the fit results.
A signal yield of $61 \pm 28$ events is obtained when evaluating the signal hypothesis in the baseline SR,
with a fitted edge position of $144.2^{+3.3}_{-2.2}\GeV$. This is in agreement with the upwards fluctuations in the mass
region between 96 and 150\GeV in the counting experiment and corresponds to a local significance of 2.3 standard deviations.
To estimate the global $p$-value~\cite{Gross:2010qma} of the result, the test statistic $-2\ln Q$,
where $Q$ denotes the ratio of the fitted likelihood value for the signal-plus-background
hypothesis to the background-only hypothesis, is evaluated on data and compared to the respective quantity on a large sample
of background-only pseudo-experiments where the edge position can have any value. The resulting $p$-value is interpreted as the one-sided tail probability
of a Gaussian distribution and corresponds to an excess in the observed number of events compared to the SM background prediction
with a global significance of 1.5 standard deviations.

\begin{table}[!hbtp]
\renewcommand{\arraystretch}{1.2}
\centering
\topcaption{Results of the unbinned maximum likelihood fit for event yields in the edge fit SR of Table~\ref{tab:selections_signalRegions},
  including the \dyjets and FS background components,
  along with the fitted signal contribution and edge position. The fitted value for \Rsfof and the local and global signal significances in terms of standard deviations
  are also given.
The uncertainties account for both statistical and systematic components.}
\label{tab:fitResults}
\begin{tabular}{l c}
\hline
  \dyjets yield           & $191 \pm 19$        \\
  FS yield                & $768 \pm 24$         \\
  \Rsfof                  & \x$1.07 \pm 0.03$              \\
  Signal yield            & \x$61 \pm 28$       \\
  $\mll^\text{edge} $      & $144.2^{+3.3}_{-2.2}\GeV$  \\
  \hline
  Local significance                   & 2.3 s.d.          \\
  Global significance                  & 1.5 s.d.          \\
\hline
\end{tabular}
\end{table}

\section{Interpretation}
\label{sec:interpretation}

The results are interpreted in terms of the simplified models defined in Section~\ref{sec:signalmodels}.
Upper limits on the cross section (assuming branching fractions presented in Section~\ref{sec:signalmodels}) have been calculated at 95\%
confidence level (CL) using the CL$_\mathrm{S}$ criterion and an asymptotic formulation~\cite{Junk:1999kv,0954-3899-28-10-313,HiggsTool1,Cowan:2010js},
taking into account the statistical and systematic uncertainties in the signal yields and the background predictions.

\subsection{Systematic uncertainty in the signal yield}

The systematic uncertainties in the signal yield are summarized in Table~\ref{tab:systs}.
The uncertainty in the measurement of the integrated luminosity is 2.5\%~\cite{CMS-PAS-LUM-17-001}.
The uncertainty in the lepton identification
and isolation efficiency amounts to 5\% in the signal acceptance.
A further uncertainty of 4\% arises from the modeling of the lepton efficiency in the fast simulation used for signal.
The uncertainties in the b tagging efficiency and mistag probability are between 0 and 5\% depending on the signal model and masses probed.
The uncertainty in the trigger efficiency is 3\%.
The uncertainty in the jet energy scale varies between 0--5\% depending on the signal kinematics.
The uncertainty associated with the modeling of initial-state radiation (ISR) is 0--2.5\%.
Determining the signal acceptance in high- and low-pileup regimes separately yields an uncertainty of 1--2\%.
The uncertainty in the \ptmiss modeling in fast simulation amounts to 0--4\%.
Generator renormalization and factorization scales are varied up and down by a factor of two, resulting
in an uncertainty in the signal acceptance of 1--3\%.
Finally the statistical uncertainty in the number of simulated events is also considered and found
to be in the range 1--15\%, depending on the SR and mass point.

\begin{table}[htb]
\renewcommand*{\arraystretch}{1.1}
\centering
\topcaption{\label{tab:systs}
Systematic uncertainties taken into account for the signal yields and their typical values.}
\begin{tabular}{l c}
\hline
Source of uncertainty                & Uncertainty (\%)     \\
\hline
Integrated luminosity                & 2.5                  \\
Lepton reconstruction and isolation  & 5                    \\
Fast simulation lepton efficiency    & 4                    \\
b tag modeling                       & 0--5                  \\
Trigger modeling                     & 3                    \\
Jet energy scale                     & 0--5                  \\
ISR modeling                         & \x\y0--2.5                 \\
Pileup                               & 1--2              	\\
Fast simulation \ptmiss modeling        & 0--4                 \\
Renorm./fact. scales                 & 1--3                   \\
Statistical uncertainty              & \x1--15                  \\
\hline
Total uncertainty                    & \x9--18                \\
\hline
\end{tabular}
\end{table}

\subsection{Interpretations using simplified models}

\begin{figure}[!hb]
 \centering
   \includegraphics[width=0.60\textwidth]{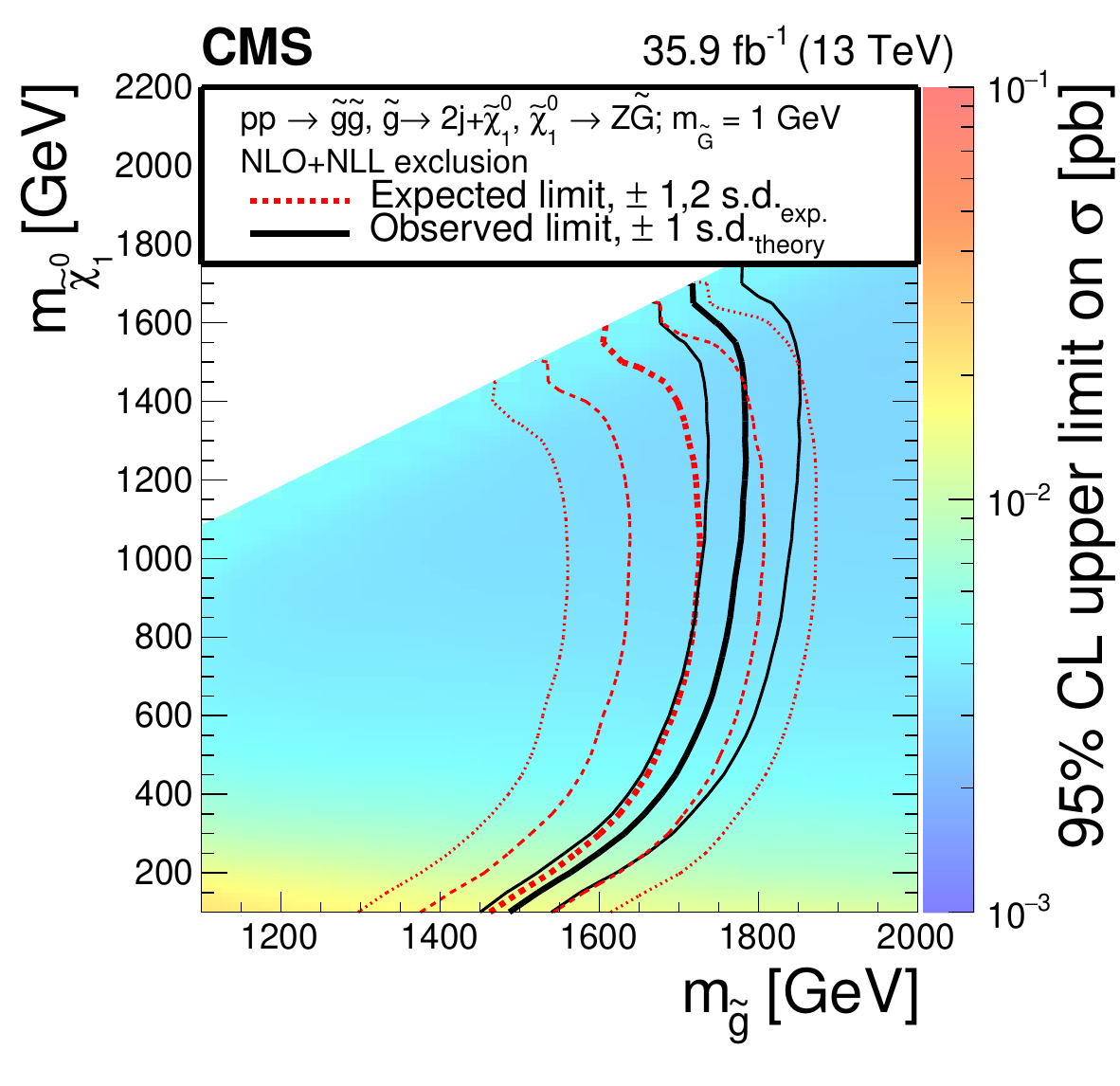}
   \caption{\label{fig:Limits1}
     Cross section upper limit and exclusion contours at 95\% CL for the gluino GMSB model as a function of the \gluino~and \firstchi masses,
     obtained from the results of the strong production on-\PZ search.
     The region to the left of the thick red dotted (black solid) line is excluded by the expected (observed) limit.
     The thin red dotted curves indicate the regions containing 68 and 95\% of the distribution of limits
     expected under the background-only hypothesis.
     The thin solid black curves show the change in the observed limit due to
     variation of the signal cross sections within their theoretical uncertainties.
   }
\end{figure}

The gluino GMSB model leads to a signature containing at least six jets in the final state when one of the \PZ bosons decays leptonically
and the other decays hadronically.  Therefore, most of the sensitivity of the
on-\PZ search is provided by the high jet multiplicity SRs.
All of the on-\PZ strong-production SRs are considered, however, to set limits in this model.
The expected and observed limits are presented in Fig.~\ref{fig:Limits1} as a function of the \gluino{}  and \firstchi masses.
We are able to probe gluino masses up to 1500--1770\GeV depending on the mass of \firstchi.
This represents an improvement of around 500\GeV compared to the previously published CMS result~\cite{CMS:Zedge2015}.

\begin{figure}[hb]
 \centering
   \includegraphics[width=0.6\textwidth]{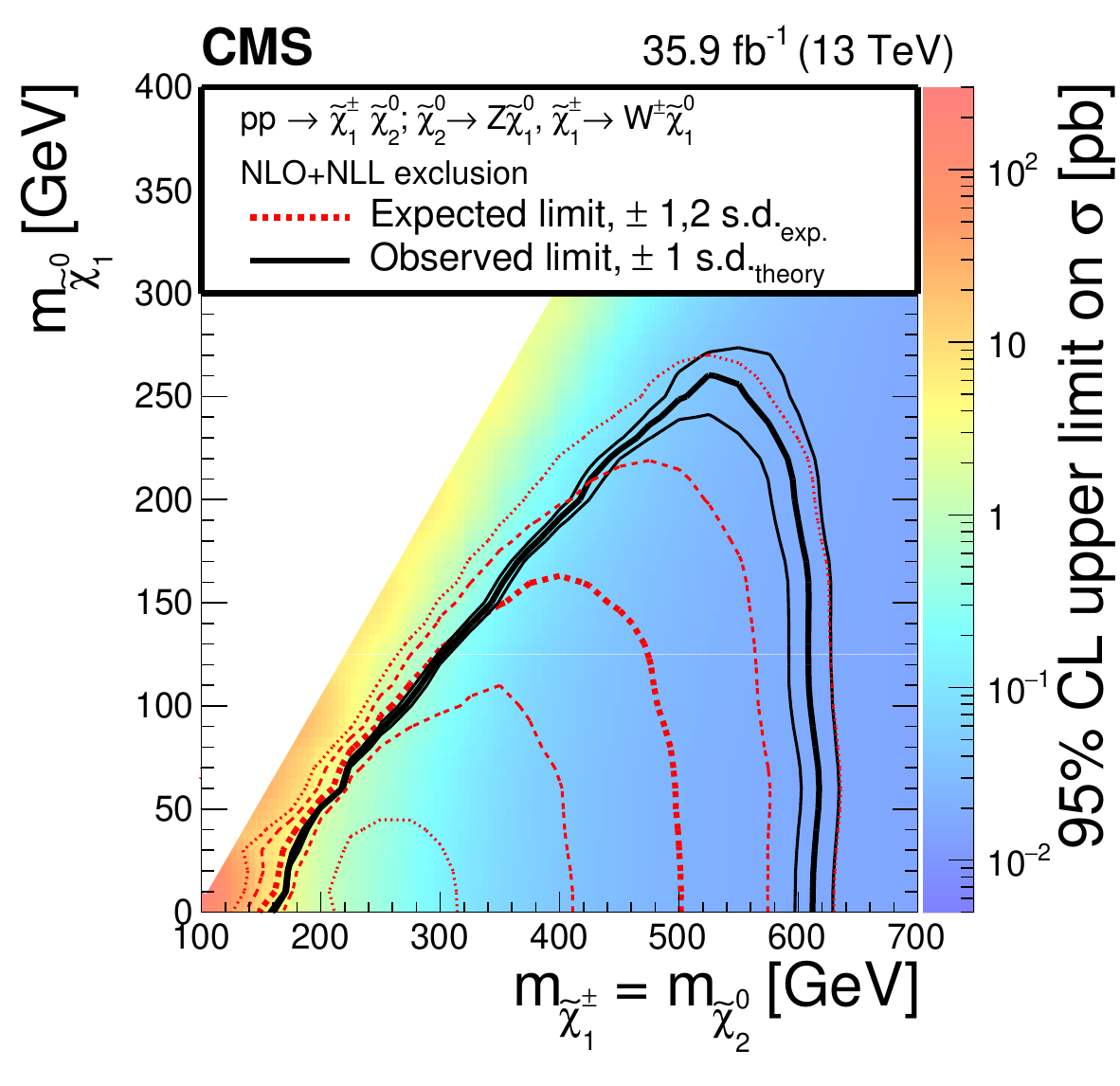}
   \caption{\label{fig:LimitTChiWZ}
     Cross section upper limit and exclusion contours at 95\% CL for the EW $\PW\PZ$ model
     as a function of the \firstcharg (equal to \secondchi) and \firstchi masses,
     obtained using the on-\PZ search for EW production results.
     The region under the thick red dotted (black solid) line is excluded by the expected (observed) limit.
     The thin red dotted curves indicate the regions containing 68 and 95\% of the distribution of limits
     expected under the background-only hypothesis.
     The thin solid black curves show the change in the observed limit due to
     variation of the signal cross sections within their theoretical uncertainties.
   }
\end{figure}

The on-\PZ search for EW production is interpreted using the models described in Section~\ref{sec:signalmodels}.
For the model of \firstcharg\secondchi production with decays to $\PW\PZ$, the \vz SR
provides almost all of the sensitivity.
Figure~\ref{fig:LimitTChiWZ} shows the cross section upper limits and the exclusion lines at 95\% CL,
as a function of the \firstcharg (or \secondchi) and \firstchi masses.
The analysis probes \firstcharg masses between approximately 160 and 610\GeV, depending on the mass of \firstchi.
The observed limit is stronger than expected due to the observed yields being smaller than predicted
in the highest two \ptmiss\ bins of the \vz SR.
This result extends the observed exclusion using 8\TeV data by around 300\GeV in the mass of \firstcharg~\cite{2012ewk}.

\begin{figure}
\centering
\includegraphics[width=0.6\textwidth]{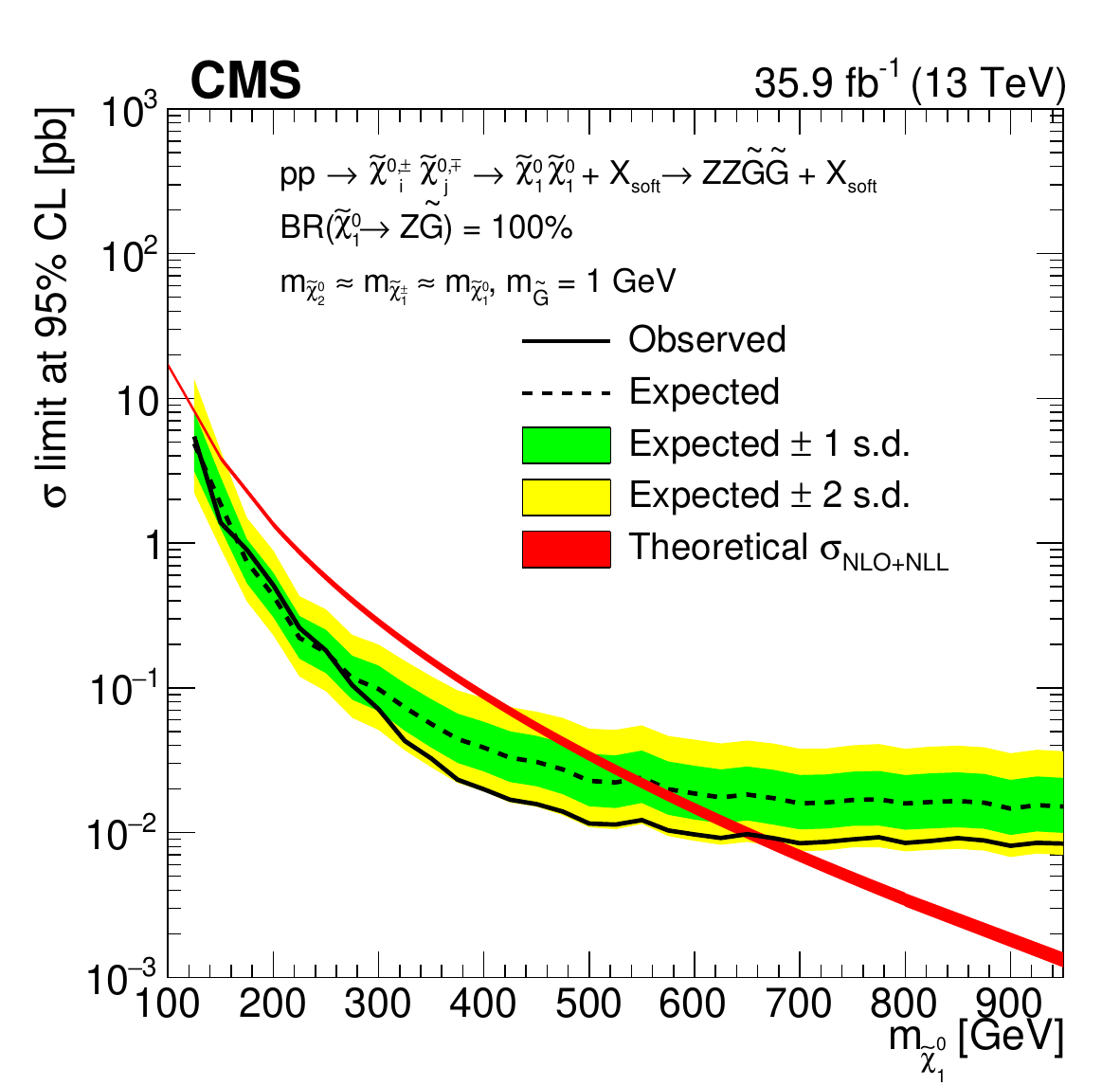}
\includegraphics[width=0.6\textwidth]{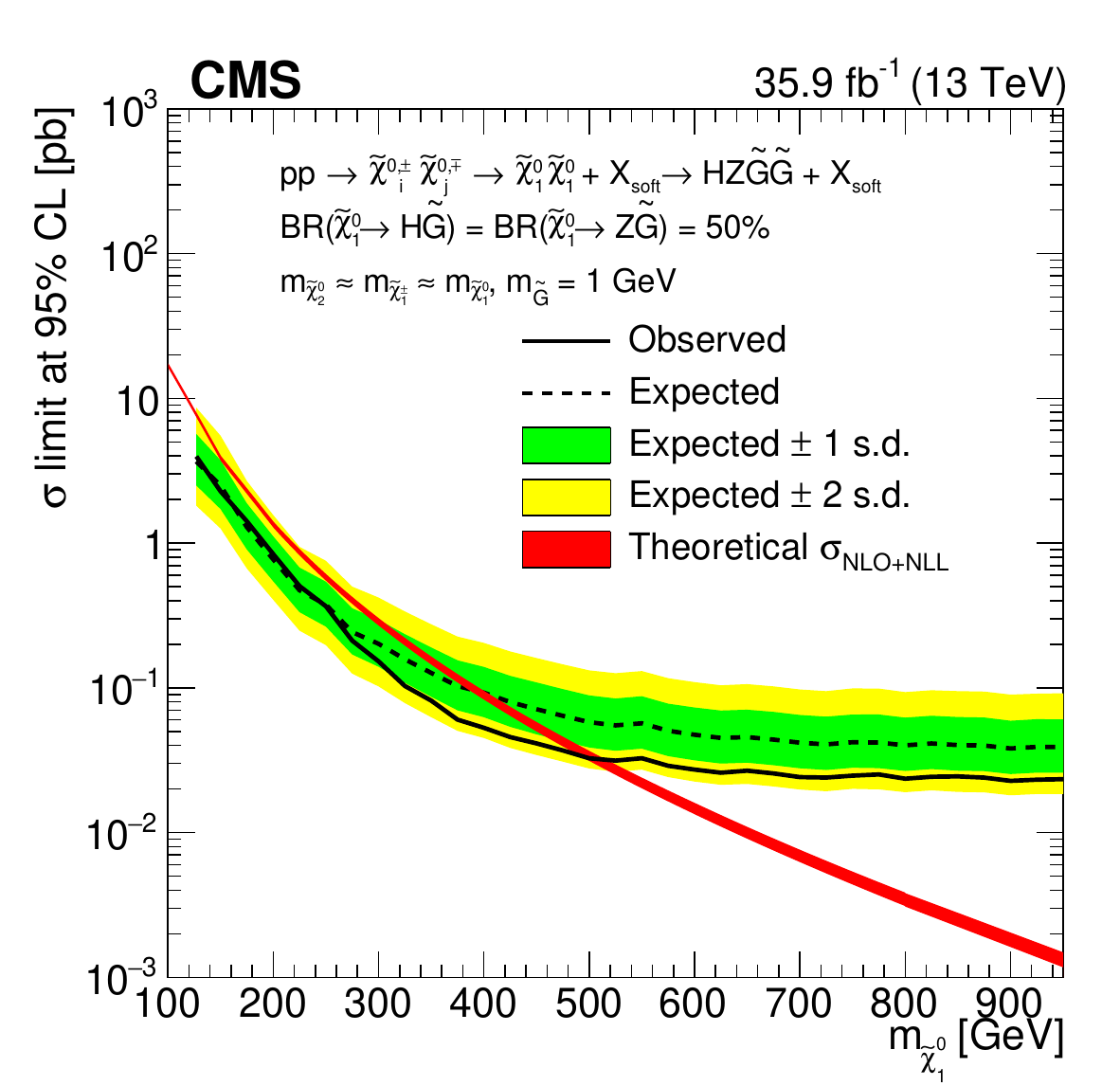}
\caption{Cross section upper limit and exclusion lines at 95\% CL, as a function of the \firstchi mass,
  for the search for EW production in the $\PZ\PZ$ topology (upper)
  and with a 50\% branching fraction to each of the \PZ and Higgs bosons (lower).
The red band shows the theoretical cross section, with the thickness of band representing the theoretical uncertainty in the signal cross section.
  Regions where the black dotted line reaches below the theoretical cross section are expected to be excluded.
     The green (yellow) band indicates the region containing 68 (95)\% of the distribution of limits
     expected under the background-only hypothesis.
     The observed upper limit on the cross section is shown with a solid black line.
}
\label{fig:LimitTChiZZHZ}
\end{figure}

For the model of \firstchi\firstchi production with decays to $\PZ\PZ$, the \vz region contains most of the signal,
but the $\PH\PZ$ SR accepts the events where the \PZ boson decays to \bbbar.
The limit is shown in Fig.~\ref{fig:LimitTChiZZHZ}~(upper)
as a function of the \firstchi mass.  We probe masses up to around 650\GeV.
The observed limit is stronger than the expected due to the deficit of observed events in the high-\ptmiss\ bins of the \vz SR.
This result extends the observed limit by around 300\GeV compared to the result using 8\TeV data~\cite{2012ewkhiggs}.

For the model of \firstchi\firstchi production with decays to $\PH\PZ$, the $\PH\PZ$ SR dominates the expected limit.
The maximal branching fraction to the $\PH\PZ$ final state is 50\%,
achieved when \firstchi decays with 50\% probability to either the \PZ or Higgs boson.
In this scenario, one also expects to have a 25\% branching fraction to the $\PZ\PZ$ topology.
We set limits on the 50\% branching fraction model in Fig.~\ref{fig:LimitTChiZZHZ}~(lower) using these assumptions
and considering the signal contributions from both the $\PZ\PZ$ and $\PH\PZ$ topologies.
In this mixed decay model, we probe masses up to around 500\GeV.
The observed limit at high masses is dominated by the same effect as in the pure $\PZ\PZ$ topology.
For masses below 200\GeV, the events from the $\PH\PZ$ topology alone give an expected exclusion that is
2--5 times more stringent than those from the $\PZ\PZ$ topology alone, while for higher masses, the two
topologies yield expected limits that are similar to within 30\%.
The previous exclusion limit using 8\TeV data is extended by around 200\GeV.

\begin{figure}
  \centering
    \includegraphics[width=0.6\textwidth]{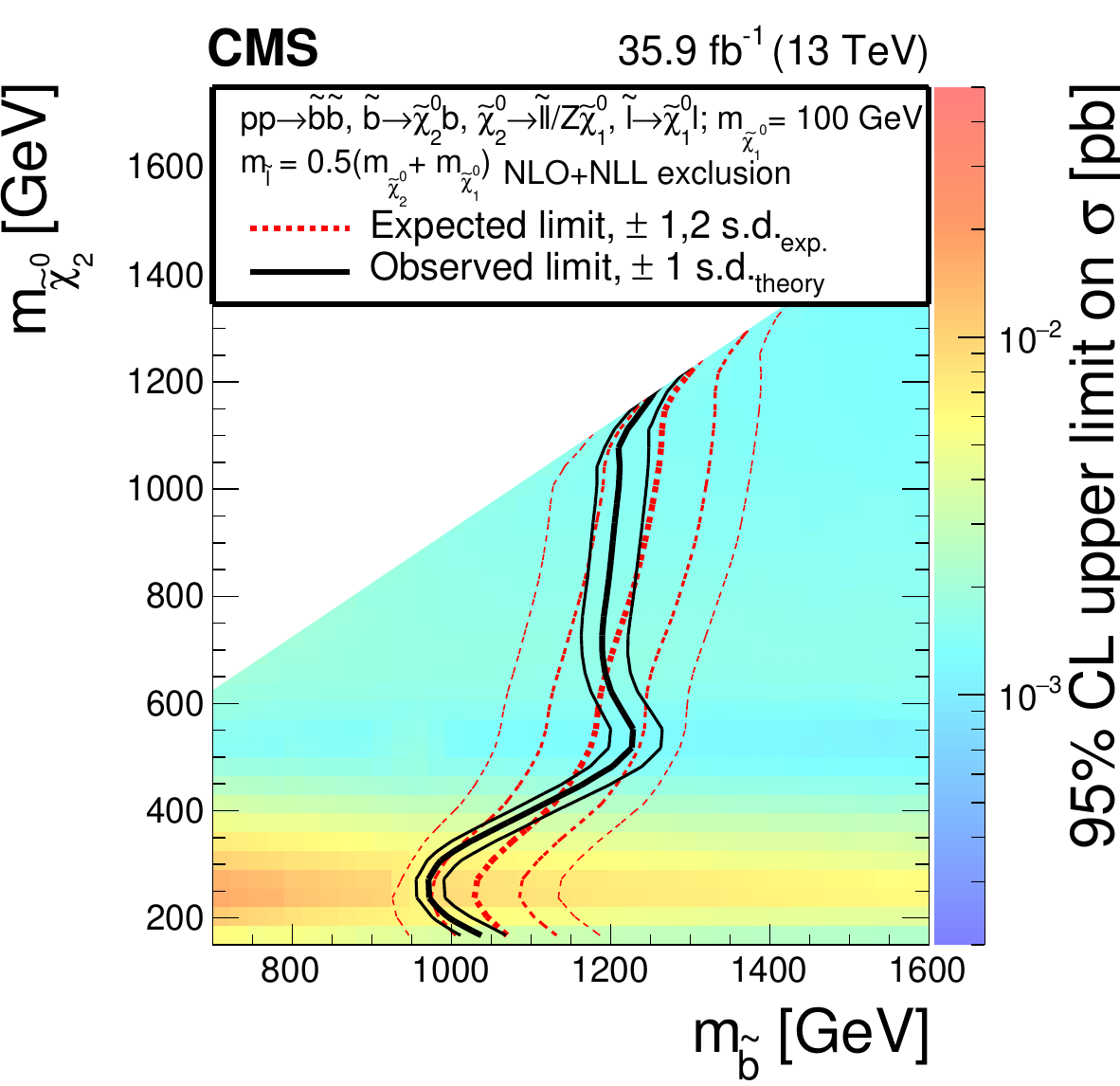}
    \caption{Cross section upper limit and exclusion contours at 95\% CL for the slepton edge model
      as a function of the \sbottom and \secondchi masses,
      obtained from the results of the edge search.
      The region to the left of the thick red dotted (black solid) line is excluded by the expected (observed) limit.
     The thin red dotted curves indicate the regions containing 68 and 95\% of the distribution of limits
     expected under the background-only hypothesis.
     The thin solid black curves show the change in the observed limit due to
     variation of the signal cross sections within their theoretical uncertainties.
    }
    \label{fig:Limits2}
\end{figure}

The edge search is interpreted using the slepton edge model, combining the seven \mll bins and the two likelihood regions.
Figure~\ref{fig:Limits2} shows
the exclusion contour as a function of the \sbottom and \secondchi masses. We exclude \sbottom masses up to around 980--1200\GeV,
depending on the mass of \secondchi, extending previous exclusion limits in the same model by 400--600\GeV.
A decrease of the sensitivity is observed for those models where the \secondchi mass is in the range 200--300\GeV.
The \mll distribution for these models has an edge in the range 100--200\GeV,
and most of the signal events fall either into the SRs with the highest background prediction
or in the range $86 < \mll < 96\GeV$, which is not considered for this part of the analysis.
The observed limit in this regime is weaker than the expected one due
to the deviation in the not-\ttbar-like, 96--150\GeV mass bin.
For high \secondchi masses, the majority of signal events fall into the highest mass bins, which are nearly background free. This results in
increased sensitivity for these mass points. In the highest not-\ttbar-like \mll bin, 5 events are observed and 1.6 are expected, yielding a
weaker observed limit for these mass points. The not-\ttbar-like \mll bin of 300--400\GeV contains 0 observed events
compared to an expectation of 3.5 events, yielding the
stronger observed limit for the \secondchi masses of about 500\GeV.

\section{Summary}
\label{sec:summary}
A search for phenomena beyond the standard model (SM) in events with opposite-charge, same-flavor leptons, jets, and missing transverse momentum
has been presented. The data used corresponds to a sample of
\Pp\Pp\ collisions collected with the CMS detector in 2016 at a center-of-mass energy of 13\TeV, corresponding to an integrated luminosity of \lint.
Searches are performed for signals with a dilepton invariant mass (\mll) compatible with the \PZ boson
or producing a kinematic edge in the distribution of \mll.
By comparing the observation to estimates for SM backgrounds obtained from data control samples, no statistically significant
evidence for a signal has been observed.

The search for strongly produced new physics containing an on-shell \PZ boson is interpreted
in a model of gauge-mediated supersymmetry breaking (GMSB),
where the \PZ bosons are produced in decay chains initiated through gluino pair production.
Gluino masses below 1500--1770\GeV have been excluded, depending on the neutralino mass,
extending the exclusion limits derived from the previous CMS publication by almost 500\GeV.

The search for electroweak production with an on-shell \PZ boson has been interpreted in multiple simplified models.
For chargino-neutralino production,
where the neutralino decays to a \PZ boson and the lightest supersymmetric particle (LSP) and the chargino decays to a \PW\ boson and the LSP,
we probe chargino masses in the range 160--610\GeV.
In a GMSB model of neutralino-neutralino production decaying to $\PZ\PZ$ and LSPs, we probe neutralino masses up to around 650\GeV.
Assuming GMSB production where the neutralino has a branching fraction of 50\% to the \PZ boson and 50\% to the Higgs boson,
we probe neutralino masses up to around 500\GeV.
Compared to published CMS results using 8\TeV data, these extend the exclusion limits
by around 200--300\GeV depending on the model.

The search for a kinematic edge in the \mll\ distribution is interpreted in a simplified model based on bottom squark pair production.
Decay chains containing the two lightest neutralinos and a slepton are assumed,
leading to edge-like signatures in the distribution of \mll.
Bottom squark masses below 980--1200\GeV
have been excluded, depending on the mass of the second neutralino.
These extend the previous CMS exclusion limits in the same model by 400--600\GeV.

\begin{acknowledgments}
We congratulate our colleagues in the CERN accelerator departments for the excellent performance of the LHC and thank the technical and administrative staffs at CERN and at other CMS institutes for their contributions to the success of the CMS effort. In addition, we gratefully acknowledge the computing centers and personnel of the Worldwide LHC Computing Grid for delivering so effectively the computing infrastructure essential to our analyses. Finally, we acknowledge the enduring support for the construction and operation of the LHC and the CMS detector provided by the following funding agencies: BMWFW and FWF (Austria); FNRS and FWO (Belgium); CNPq, CAPES, FAPERJ, and FAPESP (Brazil); MES (Bulgaria); CERN; CAS, MoST, and NSFC (China); COLCIENCIAS (Colombia); MSES and CSF (Croatia); RPF (Cyprus); SENESCYT (Ecuador); MoER, ERC IUT, and ERDF (Estonia); Academy of Finland, MEC, and HIP (Finland); CEA and CNRS/IN2P3 (France); BMBF, DFG, and HGF (Germany); GSRT (Greece); OTKA and NIH (Hungary); DAE and DST (India); IPM (Iran); SFI (Ireland); INFN (Italy); MSIP and NRF (Republic of Korea); LAS (Lithuania); MOE and UM (Malaysia); BUAP, CINVESTAV, CONACYT, LNS, SEP, and UASLP-FAI (Mexico); MBIE (New Zealand); PAEC (Pakistan); MSHE and NSC (Poland); FCT (Portugal); JINR (Dubna); MON, RosAtom, RAS, RFBR and RAEP (Russia); MESTD (Serbia); SEIDI, CPAN, PCTI and FEDER (Spain); Swiss Funding Agencies (Switzerland); MST (Taipei); ThEPCenter, IPST, STAR, and NSTDA (Thailand); TUBITAK and TAEK (Turkey); NASU and SFFR (Ukraine); STFC (United Kingdom); DOE and NSF (USA).

\hyphenation{Rachada-pisek} Individuals have received support from the Marie-Curie program and the European Research Council and Horizon 2020 Grant, contract No. 675440 (European Union); the Leventis Foundation; the A. P. Sloan Foundation; the Alexander von Humboldt Foundation; the Belgian Federal Science Policy Office; the Fonds pour la Formation \`a la Recherche dans l'Industrie et dans l'Agriculture (FRIA-Belgium); the Agentschap voor Innovatie door Wetenschap en Technologie (IWT-Belgium); the Ministry of Education, Youth and Sports (MEYS) of the Czech Republic; the Council of Science and Industrial Research, India; the HOMING PLUS program of the Foundation for Polish Science, cofinanced from European Union, Regional Development Fund, the Mobility Plus program of the Ministry of Science and Higher Education, the National Science Center (Poland), contracts Harmonia 2014/14/M/ST2/00428, Opus 2014/13/B/ST2/02543, 2014/15/B/ST2/03998, and 2015/19/B/ST2/02861, Sonata-bis 2012/07/E/ST2/01406; the National Priorities Research Program by Qatar National Research Fund; the Programa Severo Ochoa del Principado de Asturias; the Thalis and Aristeia programs cofinanced by EU-ESF and the Greek NSRF; the Rachadapisek Sompot Fund for Postdoctoral Fellowship, Chulalongkorn University and the Chulalongkorn Academic into Its 2nd Century Project Advancement Project (Thailand); the Welch Foundation, contract C-1845; and the Weston Havens Foundation (USA).
\end{acknowledgments}

\bibliography{auto_generated}

\providecommand{\href}[2]{#2}\begingroup\raggedright\begin{thebibliography}{10}%
\makeatletter
\providecommand{\hrefCMSnoop }[0]{\@secondoftwo}%
\makeatother
\providecommand{\doi}{\texttt{doi:}\begingroup \urlstyle{tt}\Url}

\bibitem{Ramond:1971gb}
\hrefCMSnoop {}{P.~Ramond, ``{Dual theory for free fermions}'',} \textit{ Phys.
  Rev. D} \textbf{ 3} (1971) 2415,
\href{http://dx.doi.org/10.1103/PhysRevD.3.2415}{\doi{10.1103/PhysRevD.3.2415}}.

\bibitem{Golfand:1971iw}
\href {http://www.jetpletters.ac.ru/ps/1584/article_24309.pdf}{{\relax Yu}.~A.
  Gol'fand and E.~P. Likhtman, ``Extension of the algebra of {P}oincar\'{e}
  group generators and violation of {P} invariance'',} \textit{ JETP Lett.}
  \textbf{ 13} (1971)
323.

\bibitem{Neveu:1971rx}
\hrefCMSnoop {}{A.~Neveu and J.~H. Schwarz, ``Factorizable dual model of
  pions'',} \textit{ Nucl. Phys. B} \textbf{ 31} (1971) 86,
\href{http://dx.doi.org/10.1016/0550-3213(71)90448-2}{\doi{10.1016/0550-3213(71)90448-2}}.

\bibitem{Volkov:1972jx}
\href {http://www.jetpletters.ac.ru/ps/1766/article_26864.pdf}{D.~V. Volkov and
  V.~P. Akulov, ``Possible universal neutrino interaction'',} \textit{ JETP
  Lett.} \textbf{ 16} (1972)
438.

\bibitem{Wess:1973kz}
\hrefCMSnoop {}{J.~Wess and B.~Zumino, ``A {L}agrangian model invariant under
  supergauge transformations'',} \textit{ Phys. Lett. B} \textbf{ 49} (1974)
  52,
\href{http://dx.doi.org/10.1016/0370-2693(74)90578-4}{\doi{10.1016/0370-2693(74)90578-4}}.

\bibitem{Wess:1974tw}
\hrefCMSnoop {}{J.~Wess and B.~Zumino, ``{Supergauge transformations in four
  dimensions}'',} \textit{ Nucl. Phys. B} \textbf{ 70} (1974) 39,
\href{http://dx.doi.org/10.1016/0550-3213(74)90355-1}{\doi{10.1016/0550-3213(74)90355-1}}.

\bibitem{Fayet:1974pd}
\hrefCMSnoop {}{P.~Fayet, ``{Supergauge invariant extension of the {H}iggs
  mechanism and a model for the electron and its neutrino}'',} \textit{ Nucl.
  Phys. B} \textbf{ 90} (1975) 104,
\href{http://dx.doi.org/10.1016/0550-3213(75)90636-7}{\doi{10.1016/0550-3213(75)90636-7}}.

\bibitem{Nilles:1983ge}
\hrefCMSnoop {}{H.~P. Nilles, ``{Supersymmetry, supergravity and particle
  physics}'',} \textit{ Phys. Rep.} \textbf{ 110} (1984) 1,
\href{http://dx.doi.org/10.1016/0370-1573(84)90008-5}{\doi{10.1016/0370-1573(84)90008-5}}.

\bibitem{Farrar:1978xj}
\hrefCMSnoop {}{G.~R. Farrar and P.~Fayet, ``Phenomenology of the production,
  decay, and detection of new hadronic states associated with supersymmetry'',}
  \textit{ Phys. Lett. B} \textbf{ 76} (1978) 575,
\href{http://dx.doi.org/10.1016/0370-2693(78)90858-4}{\doi{10.1016/0370-2693(78)90858-4}}.

\bibitem{unification}
\hrefCMSnoop {}{A.~J. Buras, J.~R. Ellis, M.~K. Gaillard, and D.~V. Nanopoulos,
  ``Aspects of the grand unification of strong, weak and electromagnetic
  interactions'',} \textit{ Nucl. Phys. B} \textbf{ 135} (1978) 66,
\href{http://dx.doi.org/10.1016/0550-3213(78)90214-6}{\doi{10.1016/0550-3213(78)90214-6}}.

\bibitem{unification2}
\hrefCMSnoop {}{H.~E. Haber and G.~L. Kane, ``The search for supersymmetry:
  Probing physics beyond the standard model'',} \textit{ Phys. Rept.} \textbf{
  117} (1985) 75,
\href{http://dx.doi.org/10.1016/0370-1573(85)90051-1}{\doi{10.1016/0370-1573(85)90051-1}}.

\bibitem{Hinchliffe:1996iu}
I.~Hinchliffe\hrefCMSnoop {}{ {et~al.}, ``Precision {SUSY} measurements at
  {CERN LHC}'',} \textit{ Phys. Rev. D} \textbf{ 55} (1997) 5520,
  \href{http://dx.doi.org/10.1103/PhysRevD.55.5520}{\doi{10.1103/PhysRevD.55.5520}},
\href{http://www.arXiv.org/abs/hep-ph/9610544}{\texttt{arXiv:hep-ph/9610544}}.

\bibitem{CMS:edge}
\hrefCMSnoop {}{{CMS Collaboration}, ``Search for physics beyond the {S}tandard
  {M}odel in events with two leptons, jets, and missing transverse momentum in
  pp collisions at {$\sqrt{s} = 8\TeV$}'',} \textit{ JHEP} \textbf{ 04} (2015)
  124,
  \href{http://dx.doi.org/10.1007/JHEP04(2015)124}{\doi{10.1007/JHEP04(2015)124}},
\href{http://www.arXiv.org/abs/1502.06031}{\texttt{arXiv:1502.06031}}.

\bibitem{CMS:Zedge2015}
\hrefCMSnoop {}{{CMS Collaboration}, ``Search for new physics in final states
  with two opposite-sign, same-flavor leptons, jets, and missing transverse
  momentum in pp collisions at {$\sqrt{s} = 13\TeV$}'',} \textit{ JHEP}
  \textbf{ 12} (2016) 013,
  \href{http://dx.doi.org/10.1007/JHEP12(2016)013}{\doi{10.1007/JHEP12(2016)013}},
\href{http://www.arXiv.org/abs/1607.00915}{\texttt{arXiv:1607.00915}}.

\bibitem{OSpaperCMS7TeV}
\hrefCMSnoop {}{{CMS Collaboration}, ``Search for new physics in events with
  opposite-sign leptons, jets, and missing transverse energy in pp collisions
  at {$\sqrt{s} = 7\TeV$}'',} \textit{ Phys. Lett. B} \textbf{ 718} (2013) 815,
  \href{http://dx.doi.org/10.1016/j.physletb.2012.11.036}{\doi{10.1016/j.physletb.2012.11.036}},
  \href{http://www.arXiv.org/abs/1206.3949}{\texttt{arXiv:1206.3949}}.

\bibitem{OSpaperCMS2011}
\hrefCMSnoop {}{{CMS Collaboration}, ``Search for physics beyond the {S}tandard
  {M}odel in opposite-sign dilepton events in pp collisions at {$\sqrt{s} =
  7\TeV$}'',} \textit{ JHEP} \textbf{ 06} (2011) 26,
  \href{http://dx.doi.org/10.1007/JHEP06(2011)026}{\doi{10.1007/JHEP06(2011)026}},
  \href{http://www.arXiv.org/abs/1103.1348}{\texttt{arXiv:1103.1348}}.

\bibitem{2012ewk}
\hrefCMSnoop {}{{CMS Collaboration}, ``Searches for electroweak production of
  charginos, neutralinos, and sleptons decaying to leptons and {W}, {Z}, and
  {H}iggs bosons in pp collisions at {8\TeV}'',} \textit{ Eur. Phys. J. C}
  \textbf{ 74} (2014) 3036,
  \href{http://dx.doi.org/10.1140/epjc/s10052-014-3036-7}{\doi{10.1140/epjc/s10052-014-3036-7}},
\href{http://www.arXiv.org/abs/1405.7570}{\texttt{arXiv:1405.7570}}.

\bibitem{2012ewkhiggs}
\hrefCMSnoop {}{{CMS Collaboration}, ``Searches for electroweak neutralino and
  chargino production in channels with {H}iggs, {Z}, and {W} bosons in pp
  collisions at {8\TeV}'',} \textit{ Phys. Rev. D} \textbf{ 90} (2014) 092007,
  \href{http://dx.doi.org/10.1103/PhysRevD.90.092007}{\doi{10.1103/PhysRevD.90.092007}},
\href{http://www.arXiv.org/abs/1409.3168}{\texttt{arXiv:1409.3168}}.

\bibitem{ATLAS:edge}
\hrefCMSnoop {}{{ATLAS Collaboration}, ``Search for supersymmetry in events
  containing a same-flavour opposite-sign dilepton pair, jets, and large
  missing transverse momentum in {$\sqrt{s} = 8\TeV$} pp collisions with the
  {ATLAS} detector'',} \textit{ Eur. Phys. J. C} \textbf{ 75} (2015) 318,
  \href{http://dx.doi.org/10.1140/epjc/s10052-015-3661-9}{\doi{10.1140/epjc/s10052-015-3661-9}},
  \href{http://www.arXiv.org/abs/1503.03290}{\texttt{arXiv:1503.03290}}.
[Erratum: \emph{Eur. Phys. J. C} \textbf{75} (2015) 463].

\bibitem{ATLASewk8tev}
\hrefCMSnoop {}{{ATLAS Collaboration}, ``Search for the electroweak production
  of supersymmetric particles in {$\sqrt{s} = 8\TeV$} pp collisions with the
  {ATLAS} detector'',} \textit{ Phys. Rev. D} \textbf{ 93} (2016) 052002,
  \href{http://dx.doi.org/10.1103/PhysRevD.93.052002}{\doi{10.1103/PhysRevD.93.052002}},
\href{http://www.arXiv.org/abs/1509.07152}{\texttt{arXiv:1509.07152}}.

\bibitem{ATLASOS13tev}
\hrefCMSnoop {}{{ATLAS Collaboration}, ``Search for new phenomena in events
  containing a same-flavour opposite-sign dilepton pair, jets, and large
  missing transverse momentum in {$\sqrt{s} = 13\TeV$} pp collisions with the
  {ATLAS} detector'',} \textit{ Eur. Phys. J. C} \textbf{ 77} (2017) 144,
  \href{http://dx.doi.org/10.1140/epjc/s10052-017-4700-5}{\doi{10.1140/epjc/s10052-017-4700-5}},
\href{http://www.arXiv.org/abs/1611.05791}{\texttt{arXiv:1611.05791}}.

\bibitem{bib-sms-1}
N.~Arkani-Hamed\hrefCMSnoop {}{ {et~al.}, ``{MARMOSET}: The path from {LHC}
  data to the new {S}tandard {M}odel via on-shell effective theories'',}
  (2007).
\href{http://www.arXiv.org/abs/hep-ph/0703088}{\texttt{arXiv:hep-ph/0703088}}.

\bibitem{bib-sms-2}
\hrefCMSnoop {}{J.~Alwall, P.~Schuster, and N.~Toro, ``Simplified models for a
  first characterization of new physics at the {LHC}'',} \textit{ Phys. Rev. D}
  \textbf{ 79} (2009) 075020,
  \href{http://dx.doi.org/10.1103/PhysRevD.79.075020}{\doi{10.1103/PhysRevD.79.075020}},
\href{http://www.arXiv.org/abs/0810.3921}{\texttt{arXiv:0810.3921}}.

\bibitem{bib-sms-3}
\hrefCMSnoop {}{J.~Alwall, M.-P. Le, M.~Lisanti, and J.~G. Wacker,
  ``{Model-independent jets plus missing energy searches}'',} \textit{ Phys.
  Rev. D} \textbf{ 79} (2009) 015005,
  \href{http://dx.doi.org/10.1103/PhysRevD.79.015005}{\doi{10.1103/PhysRevD.79.015005}},
\href{http://www.arXiv.org/abs/0809.3264}{\texttt{arXiv:0809.3264}}.

\bibitem{bib-sms-4}
D.~Alves\hrefCMSnoop {}{ {et~al.}, ``Simplified models for {LHC} new physics
  searches'',} \textit{ J. Phys. G} \textbf{ 39} (2012) 105005,
  \href{http://dx.doi.org/10.1088/0954-3899/39/10/105005}{\doi{10.1088/0954-3899/39/10/105005}},
\href{http://www.arXiv.org/abs/1105.2838}{\texttt{arXiv:1105.2838}}.

\bibitem{Chatrchyan:2013sza}
\hrefCMSnoop {}{{CMS Collaboration}, ``{Interpretation of searches for
  supersymmetry with simplified models}'',} \textit{ Phys. Rev. D} \textbf{ 88}
  (2013) 052017,
  \href{http://dx.doi.org/10.1103/PhysRevD.88.052017}{\doi{10.1103/PhysRevD.88.052017}},
\href{http://www.arXiv.org/abs/1301.2175}{\texttt{arXiv:1301.2175}}.

\bibitem{Matchev:1999ft}
\hrefCMSnoop {}{K.~T. Matchev and S.~D. Thomas, ``{Higgs and Z boson signatures
  of supersymmetry}'',} \textit{ Phys. Rev. D} \textbf{ 62} (2000) 077702,
  \href{http://dx.doi.org/10.1103/PhysRevD.62.077702}{\doi{10.1103/PhysRevD.62.077702}},
\href{http://www.arXiv.org/abs/hep-ph/9908482}{\texttt{arXiv:hep-ph/9908482}}.

\bibitem{Meade:2009qv}
\hrefCMSnoop {}{P.~Meade, M.~Reece, and D.~Shih, ``{Prompt decays of general
  neutralino NLSPs at the Tevatron}'',} \textit{ JHEP} \textbf{ 05} (2010) 105,
  \href{http://dx.doi.org/10.1007/JHEP05(2010)105}{\doi{10.1007/JHEP05(2010)105}},
\href{http://www.arXiv.org/abs/0911.4130}{\texttt{arXiv:0911.4130}}.

\bibitem{Ruderman}
\hrefCMSnoop {}{J.~T. Ruderman and D.~Shih, ``{General neutralino NLSPs at the
  early LHC}'',} \textit{ JHEP} \textbf{ 08} (2012) 159,
  \href{http://dx.doi.org/10.1007/JHEP08(2012)159}{\doi{10.1007/JHEP08(2012)159}},
\href{http://www.arXiv.org/abs/1103.6083}{\texttt{arXiv:1103.6083}}.

\bibitem{Skands:2003cj}
\hrefCMSnoop {}{P.~Z. Skands {et~al.}, ``{SUSY Les Houches accord: interfacing
  SUSY spectrum calculators, decay packages, and event generators}'',} \textit{
  JHEP} \textbf{ 07} (2004) 036,
  \href{http://dx.doi.org/10.1088/1126-6708/2004/07/036}{\doi{10.1088/1126-6708/2004/07/036}},
\href{http://www.arXiv.org/abs/hep-ph/0311123}{\texttt{arXiv:hep-ph/0311123}}.

\bibitem{Chatrchyan:2008zzk}
\hrefCMSnoop {}{{CMS Collaboration}, ``The {CMS} experiment at the {CERN}
  {LHC}'',} \textit{ JINST} \textbf{ 3} (2008) S08004,
\href{http://dx.doi.org/10.1088/1748-0221/3/08/S08004}{\doi{10.1088/1748-0221/3/08/S08004}}.

\bibitem{Sirunyan:2017ulk}
\hrefCMSnoop {}{{CMS Collaboration}, ``{Particle-flow reconstruction and global
  event description with the CMS detector}'',} \textit{ JINST} \textbf{ 12}
  (2017) P10003,
  \href{http://dx.doi.org/10.1088/1748-0221/12/10/P10003}{\doi{10.1088/1748-0221/12/10/P10003}},
\href{http://www.arXiv.org/abs/1706.04965}{\texttt{arXiv:1706.04965}}.

\bibitem{Cacciari:2008gp}
\hrefCMSnoop {}{M.~Cacciari, G.~P. Salam, and G.~Soyez, ``The anti-$k_t$ jet
  clustering algorithm'',} \textit{ JHEP} \textbf{ 04} (2008) 063,
  \href{http://dx.doi.org/10.1088/1126-6708/2008/04/063}{\doi{10.1088/1126-6708/2008/04/063}},
  \href{http://www.arXiv.org/abs/0802.1189}{\texttt{arXiv:0802.1189}}.

\bibitem{FastJet}
\hrefCMSnoop {}{M.~Cacciari, G.~P. Salam, and G.~Soyez, ``{FastJet} user
  manual'',} \textit{ Eur. Phys. J. C} \textbf{ 72} (2012) 1896,
  \href{http://dx.doi.org/10.1140/epjc/s10052-012-1896-2}{\doi{10.1140/epjc/s10052-012-1896-2}},
\href{http://www.arXiv.org/abs/1111.6097}{\texttt{arXiv:1111.6097}}.

\bibitem{Khachatryan:2015hwa}
\hrefCMSnoop {}{{CMS Collaboration}, ``{Performance of electron reconstruction
  and selection with the CMS detector in proton-proton collisions at $\sqrt{s}
  = 8\TeV$}'',} \textit{ JINST} \textbf{ 10} (2015) P06005,
  \href{http://dx.doi.org/10.1088/1748-0221/10/06/P06005}{\doi{10.1088/1748-0221/10/06/P06005}},
\href{http://www.arXiv.org/abs/1502.02701}{\texttt{arXiv:1502.02701}}.

\bibitem{Rehermann:2010vq}
\hrefCMSnoop {}{K.~Rehermann and B.~Tweedie, ``Efficient identification of
  boosted semileptonic top quarks at the {LHC}'',} \textit{ JHEP} \textbf{ 03}
  (2011) 059,
  \href{http://dx.doi.org/10.1007/JHEP03(2011)059}{\doi{10.1007/JHEP03(2011)059}},
\href{http://www.arXiv.org/abs/1007.2221}{\texttt{arXiv:1007.2221}}.

\bibitem{CMSPhotonID}
\hrefCMSnoop {}{{CMS Collaboration}, ``Performance of photon reconstruction and
  identification with the {CMS} detector in proton-proton collisions at
  {$\sqrt{s} = 8\TeV$}'',} \textit{ JINST} \textbf{ 10} (2015) P08010,
  \href{http://dx.doi.org/10.1088/1748-0221/10/08/P08010}{\doi{10.1088/1748-0221/10/08/P08010}},
\href{http://www.arXiv.org/abs/1502.02702}{\texttt{arXiv:1502.02702}}.

\bibitem{Cacciari:2005hq}
\hrefCMSnoop {}{M.~Cacciari and G.~P. Salam, ``{Dispelling the N$^3$ myth for
  the $k_t$ jet-finder}'',} \textit{ Phys. Lett. B} \textbf{ 641} (2006) 57,
  \href{http://dx.doi.org/10.1016/j.physletb.2006.08.037}{\doi{10.1016/j.physletb.2006.08.037}},
  \href{http://www.arXiv.org/abs/hep-ph/0512210}{\texttt{arXiv:hep-ph/0512210}}.

\bibitem{1748-0221-6-11-P11002}
\hrefCMSnoop {}{{CMS Collaboration}, ``Determination of jet energy calibration
  and transverse momentum resolution in {CMS}'',} \textit{ JINST} \textbf{ 6}
  (2011) P11002,
  \href{http://dx.doi.org/10.1088/1748-0221/6/11/P11002}{\doi{10.1088/1748-0221/6/11/P11002}},
\href{http://www.arXiv.org/abs/1107.4277}{\texttt{arXiv:1107.4277}}.

\bibitem{cacciari-2008-659}
\hrefCMSnoop {}{M.~Cacciari and G.~P. Salam, ``Pileup subtraction using jet
  areas'',} \textit{ Phys. Lett. B} \textbf{ 659} (2008) 119,
  \href{http://dx.doi.org/10.1016/j.physletb.2007.09.077}{\doi{10.1016/j.physletb.2007.09.077}},
  \href{http://www.arXiv.org/abs/0707.1378}{\texttt{arXiv:0707.1378}}.

\bibitem{BTV-16-002}
\hrefCMSnoop {}{{CMS Collaboration}, ``{Identification of heavy-flavour jets
  with the CMS detector in pp collisions at 13 TeV}'',} (2017).
  \href{http://www.arXiv.org/abs/1712.07158}{\texttt{arXiv:1712.07158}}.
Submitted to {\it JINST}.

\bibitem{Alwall:2014hca}
J.~Alwall\hrefCMSnoop {}{ {et~al.}, ``{The automated computation of tree-level
  and next-to-leading order differential cross sections, and their matching to
  parton shower simulations}'',} \textit{ JHEP} \textbf{ 07} (2014) 079,
  \href{http://dx.doi.org/10.1007/JHEP07(2014)079}{\doi{10.1007/JHEP07(2014)079}},
\href{http://www.arXiv.org/abs/1405.0301}{\texttt{arXiv:1405.0301}}.

\bibitem{Alioli:2009je}
\hrefCMSnoop {}{S.~Alioli, P.~Nason, C.~Oleari, and E.~Re, ``{NLO single-top
  production matched with shower in POWHEG: $s$- and $t$-channel
  contributions}'',} \textit{ JHEP} \textbf{ 09} (2009) 111,
  \href{http://dx.doi.org/10.1088/1126-6708/2009/09/111}{\doi{10.1088/1126-6708/2009/09/111}},
  \href{http://www.arXiv.org/abs/0907.4076}{\texttt{arXiv:0907.4076}}.
[Erratum: \DOI{10.1007/JHEP02(2010)011}].

\bibitem{Re:2010bp}
\hrefCMSnoop {}{E.~Re, ``{Single-top Wt-channel production matched with parton
  showers using the POWHEG method}'',} \textit{ Eur. Phys. J. C} \textbf{ 71}
  (2011) 1547,
  \href{http://dx.doi.org/10.1140/epjc/s10052-011-1547-z}{\doi{10.1140/epjc/s10052-011-1547-z}},
\href{http://www.arXiv.org/abs/1009.2450}{\texttt{arXiv:1009.2450}}.

\bibitem{Gavin:2010az}
\hrefCMSnoop {}{R.~Gavin, Y.~Li, F.~Petriello, and S.~Quackenbush, ``{FEWZ 2.0:
  A code for hadronic Z production at next-to-next-to-leading order}'',}
  \textit{ Comput. Phys. Commun.} \textbf{ 182} (2011) 2388,
  \href{http://dx.doi.org/10.1016/j.cpc.2011.06.008}{\doi{10.1016/j.cpc.2011.06.008}},
\href{http://www.arXiv.org/abs/1011.3540}{\texttt{arXiv:1011.3540}}.

\bibitem{Gavin:2012sy}
\hrefCMSnoop {}{R.~Gavin, Y.~Li, F.~Petriello, and S.~Quackenbush, ``{W}
  physics at the {LHC} with {FEWZ 2.1}'',} \textit{ Comput. Phys. Commun.}
  \textbf{ 184} (2013) 208,
  \href{http://dx.doi.org/10.1016/j.cpc.2012.09.005}{\doi{10.1016/j.cpc.2012.09.005}},
\href{http://www.arXiv.org/abs/1201.5896}{\texttt{arXiv:1201.5896}}.

\bibitem{Czakon:2011xx}
\hrefCMSnoop {}{M.~Czakon and A.~Mitov, ``{Top++}: a program for the
  calculation of the top-pair cross-section at hadron colliders'',} \textit{
  Comput. Phys. Commun.} \textbf{ 185} (2014) 2930,
  \href{http://dx.doi.org/10.1016/j.cpc.2014.06.021}{\doi{10.1016/j.cpc.2014.06.021}},
\href{http://www.arXiv.org/abs/1112.5675}{\texttt{arXiv:1112.5675}}.

\bibitem{Borschensky:2014cia}
C.~Borschensky\hrefCMSnoop {}{ {et~al.}, ``{Squark and gluino production cross
  sections in pp collisions at $\sqrt{s} = $ 13, 14, 33 and 100\TeV}'',}
  \textit{ Eur. Phys. J. C} \textbf{ 74} (2014) 3174,
  \href{http://dx.doi.org/10.1140/epjc/s10052-014-3174-y}{\doi{10.1140/epjc/s10052-014-3174-y}},
\href{http://www.arXiv.org/abs/1407.5066}{\texttt{arXiv:1407.5066}}.

\bibitem{Fuks:2012qx}
\hrefCMSnoop {}{B.~Fuks, M.~Klasen, D.~R. Lamprea, and M.~Rothering, ``{Gaugino
  production in proton-proton collisions at a center-of-mass energy of
  8\TeV}'',} \textit{ JHEP} \textbf{ 10} (2012) 081,
  \href{http://dx.doi.org/10.1007/JHEP10(2012)081}{\doi{10.1007/JHEP10(2012)081}},
\href{http://www.arXiv.org/abs/1207.2159}{\texttt{arXiv:1207.2159}}.

\bibitem{Fuks:2013vua}
\hrefCMSnoop {}{B.~Fuks, M.~Klasen, D.~R. Lamprea, and M.~Rothering,
  ``{Precision predictions for electroweak superpartner production at hadron
  colliders with {\sc Resummino}}'',} \textit{ Eur. Phys. J. C} \textbf{ 73}
  (2013) 2480,
  \href{http://dx.doi.org/10.1140/epjc/s10052-013-2480-0}{\doi{10.1140/epjc/s10052-013-2480-0}},
\href{http://www.arXiv.org/abs/1304.0790}{\texttt{arXiv:1304.0790}}.

\bibitem{Alwall:2007fs}
\hrefCMSnoop {}{J.~Alwall {et~al.}, ``{Comparative study of various algorithms
  for the merging of parton showers and matrix elements in hadronic
  collisions}'',} \textit{ Eur. Phys. J. C} \textbf{ 53} (2008) 473,
  \href{http://dx.doi.org/10.1140/epjc/s10052-007-0490-5}{\doi{10.1140/epjc/s10052-007-0490-5}},
\href{http://www.arXiv.org/abs/0706.2569}{\texttt{arXiv:0706.2569}}.

\bibitem{powheg}
\hrefCMSnoop {}{S.~Frixione, P.~Nason, and C.~Oleari, ``Matching {NLO QCD}
  computations with parton shower simulations: the {POWHEG} method'',} \textit{
  JHEP} \textbf{ 11} (2007) 070,
  \href{http://dx.doi.org/10.1088/1126-6708/2007/11/070}{\doi{10.1088/1126-6708/2007/11/070}},
\href{http://www.arXiv.org/abs/0709.2092}{\texttt{arXiv:0709.2092}}.

\bibitem{Sjostrand:2007gs}
\hrefCMSnoop {}{T.~Sj{\"o}strand, S.~Mrenna, and P.~Z. Skands, ``A brief
  introduction to {PYTHIA 8.1}'',} \textit{ Comput. Phys. Commun.} \textbf{
  178} (2008) 852,
  \href{http://dx.doi.org/10.1016/j.cpc.2008.01.036}{\doi{10.1016/j.cpc.2008.01.036}},
\href{http://www.arXiv.org/abs/0710.3820}{\texttt{arXiv:0710.3820}}.

\bibitem{Ball:2014uwa}
\hrefCMSnoop {}{{NNPDF} Collaboration, ``{Parton distributions for the LHC Run
  II}'',} \textit{ JHEP} \textbf{ 04} (2015) 040,
  \href{http://dx.doi.org/10.1007/JHEP04(2015)040}{\doi{10.1007/JHEP04(2015)040}},
\href{http://www.arXiv.org/abs/1410.8849}{\texttt{arXiv:1410.8849}}.

\bibitem{Geant}
\hrefCMSnoop {}{{GEANT4} Collaboration, ``{GEANT4} --- a simulation toolkit'',}
  \textit{ Nucl. Instrum. Meth. A} \textbf{ 506} (2003) 250,
  \href{http://dx.doi.org/10.1016/S0168-9002(03)01368-8}{\doi{10.1016/S0168-9002(03)01368-8}}.

\bibitem{fastsim}
S.~Abdullin\hrefCMSnoop {}{ {et~al.}, ``The fast simulation of the {CMS}
  detector at {LHC}'',} \textit{ J. Phys. Conf. Ser.} \textbf{ 331} (2011)
  032049,
\href{http://dx.doi.org/10.1088/1742-6596/331/3/032049}{\doi{10.1088/1742-6596/331/3/032049}}.

\bibitem{MT2variable}
\hrefCMSnoop {}{C.~G. Lester and D.~J. Summers, ``Measuring masses of
  semiinvisibly decaying particles pair produced at hadron colliders'',}
  \textit{ Phys. Lett. B} \textbf{ 463} (1999) 99,
  \href{http://dx.doi.org/10.1016/S0370-2693(99)00945-4}{\doi{10.1016/S0370-2693(99)00945-4}},
\href{http://www.arXiv.org/abs/hep-ph/9906349}{\texttt{arXiv:hep-ph/9906349}}.

\bibitem{MT2variable2}
\hrefCMSnoop {}{A.~Barr, C.~Lester, and P.~Stephens, ``{A variable for
  measuring masses at hadron colliders when missing energy is expected;
  $M_{T2}$: the truth behind the glamour}'',} \textit{ J. Phys. G} \textbf{ 29}
  (2003) 2343,
  \href{http://dx.doi.org/10.1088/0954-3899/29/10/304}{\doi{10.1088/0954-3899/29/10/304}},
\href{http://www.arXiv.org/abs/hep-ph/0304226}{\texttt{arXiv:hep-ph/0304226}}.

\bibitem{Crystal}
\href {http://www.slac.stanford.edu/pubs/slacreports/slac-r-236.html}{M.~J.
  Oreglia, ``A study of the reactions $\psi^\prime \to \gamma \gamma \psi$''}.
\newblock PhD thesis, Stanford University, 1980.
\newblock {SLAC} Report {SLAC-R-236}, see Appendix {D}.

\bibitem{Olive:2016xmw}
\hrefCMSnoop {}{{Particle Data Group}, C.~Patrignani {et~al.}, ``{Review of
  Particle Physics}'',} \textit{ Chin. Phys. C} \textbf{ 40} (2016), no.~10,
  100001,
\href{http://dx.doi.org/10.1088/1674-1137/40/10/100001}{\doi{10.1088/1674-1137/40/10/100001}}.

\bibitem{Gross:2010qma}
\hrefCMSnoop {}{E.~Gross and O.~Vitells, ``{Trial factors or the look elsewhere
  effect in high energy physics}'',} \textit{ Eur. Phys. J. C} \textbf{ 70}
  (2010) 525,
  \href{http://dx.doi.org/10.1140/epjc/s10052-010-1470-8}{\doi{10.1140/epjc/s10052-010-1470-8}},
\href{http://www.arXiv.org/abs/1005.1891}{\texttt{arXiv:1005.1891}}.

\bibitem{Junk:1999kv}
\hrefCMSnoop {}{T.~Junk, ``{Confidence level computation for combining searches
  with small statistics}'',} \textit{ Nucl. Instrum. Meth. A} \textbf{ 434}
  (1999) 435,
  \href{http://dx.doi.org/10.1016/S0168-9002(99)00498-2}{\doi{10.1016/S0168-9002(99)00498-2}},
\href{http://www.arXiv.org/abs/hep-ex/9902006}{\texttt{arXiv:hep-ex/9902006}}.

\bibitem{0954-3899-28-10-313}
\hrefCMSnoop {}{A.~L. Read, ``{Presentation of search results: the {$\rm CL_s$}
  technique}'',} \textit{ J. Phys. G} \textbf{ 28} (2002) 2693,
\href{http://dx.doi.org/10.1088/0954-3899/28/10/313}{\doi{10.1088/0954-3899/28/10/313}}.

\bibitem{HiggsTool1}
\href {https://cds.cern.ch/record/1379837}{{ATLAS and CMS Collaborations},
  ``{Procedure for the LHC Higgs boson search combination in summer 2011}'',}
  Technical Report ATL-PHYS-PUB-2011-011, CMS-NOTE-2011-005, CERN, 2011.

\bibitem{Cowan:2010js}
\hrefCMSnoop {}{G.~Cowan, K.~Cranmer, E.~Gross, and O.~Vitells, ``Asymptotic
  formulae for likelihood-based tests of new physics'',} \textit{ Eur. Phys. J.
  C} \textbf{ 71} (2011) 1554,
  \href{http://dx.doi.org/10.1140/epjc/s10052-011-1554-0}{\doi{10.1140/epjc/s10052-011-1554-0}},
  \href{http://www.arXiv.org/abs/1007.1727}{\texttt{arXiv:1007.1727}}.
[Erratum: \DOI{10.1140/epjc/s10052-013-2501-z}].

\bibitem{CMS-PAS-LUM-17-001}
\href {http://cds.cern.ch/record/2257069}{{CMS Collaboration}, ``{CMS
  Luminosity Measurements for the 2016 Data Taking Period}'',} CMS Physics
  Analysis Summary CMS-PAS-LUM-17-001, 2017.

\bibitem{Collaboration:2242860}
\href {https://cds.cern.ch/record/2242860}{{CMS Collaboration}, ``Simplified
  likelihood for the re-interpretation of public {CMS} results'',} Technical
  Report CMS-NOTE-2017-001, 2017.

\end{thebibliography}\endgroup
\appendix
\section{Correlation and covariance matrices for the background predictions}
\label{sec:appA}

\begin{figure}[htbp]
\centering
\includegraphics[width=0.8\linewidth]{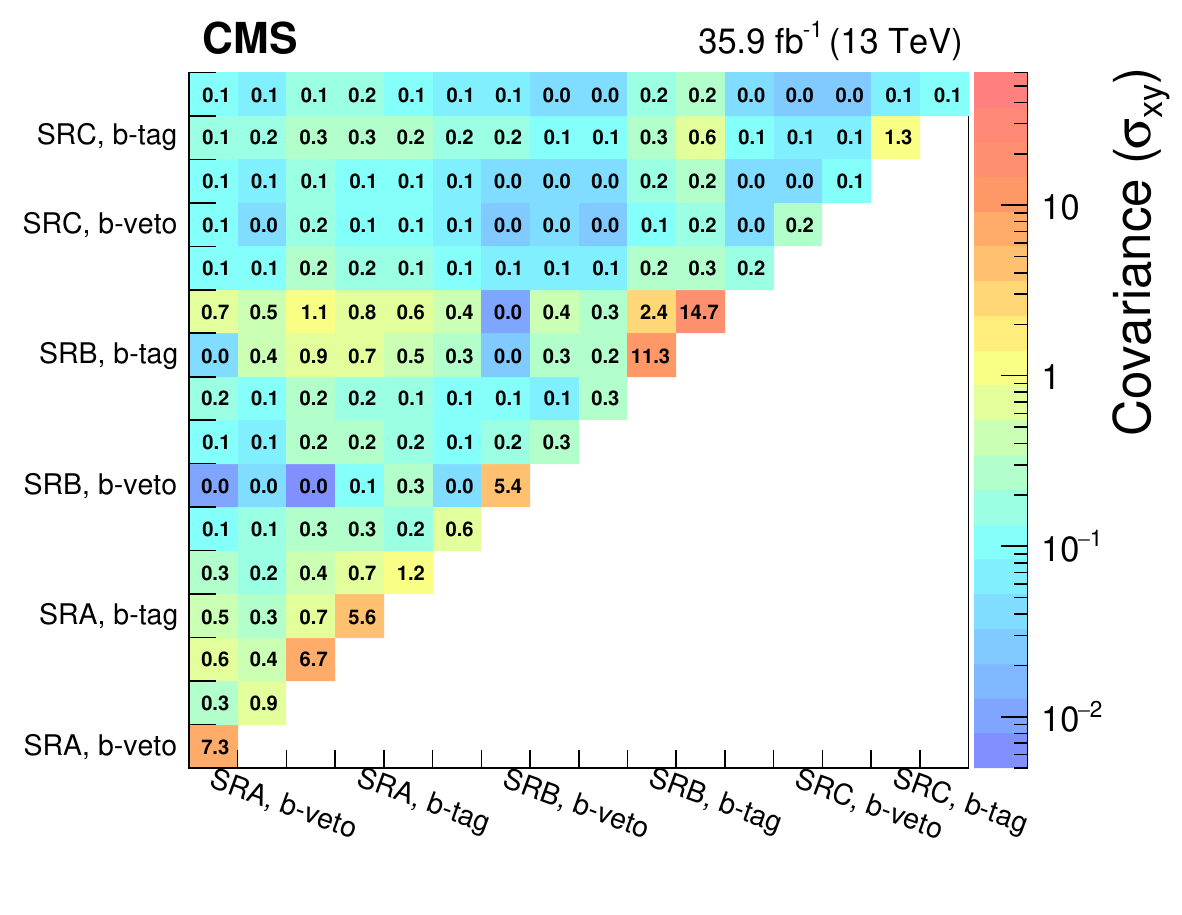}
\includegraphics[width=0.8\linewidth]{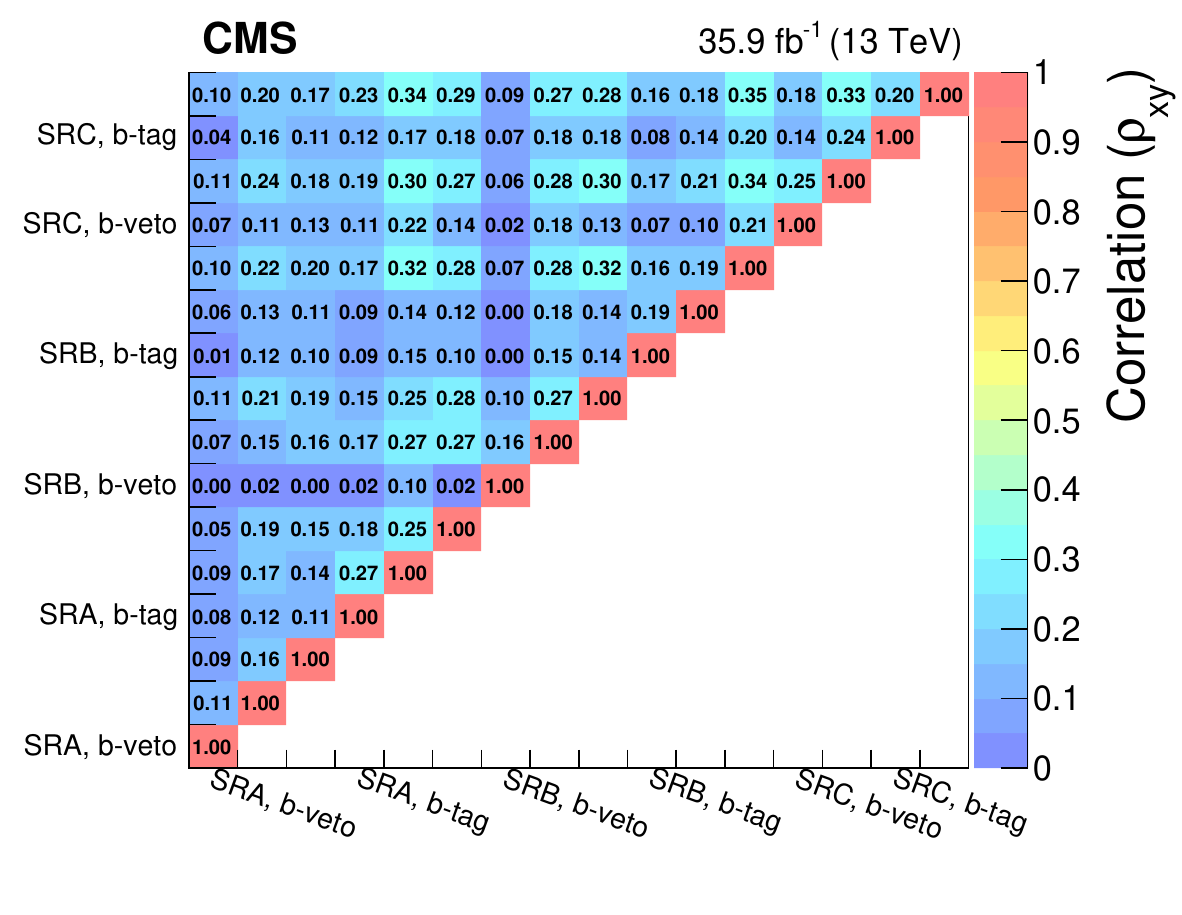}
\caption{\label{fig:correlationA}
The covariance (upper) and correlation (lower) matrices for the background predictions in the on-\PZ strong-production SRs.
Within each SR, the individual \ptmiss bins are shown in increasing order starting from 100\GeV.
The matrices are symmetric, but only the entries along and above the diagonal are shown for simplicity.
}
\end{figure}

\begin{figure}[htbp]
\centering
\includegraphics[width=0.8\linewidth]{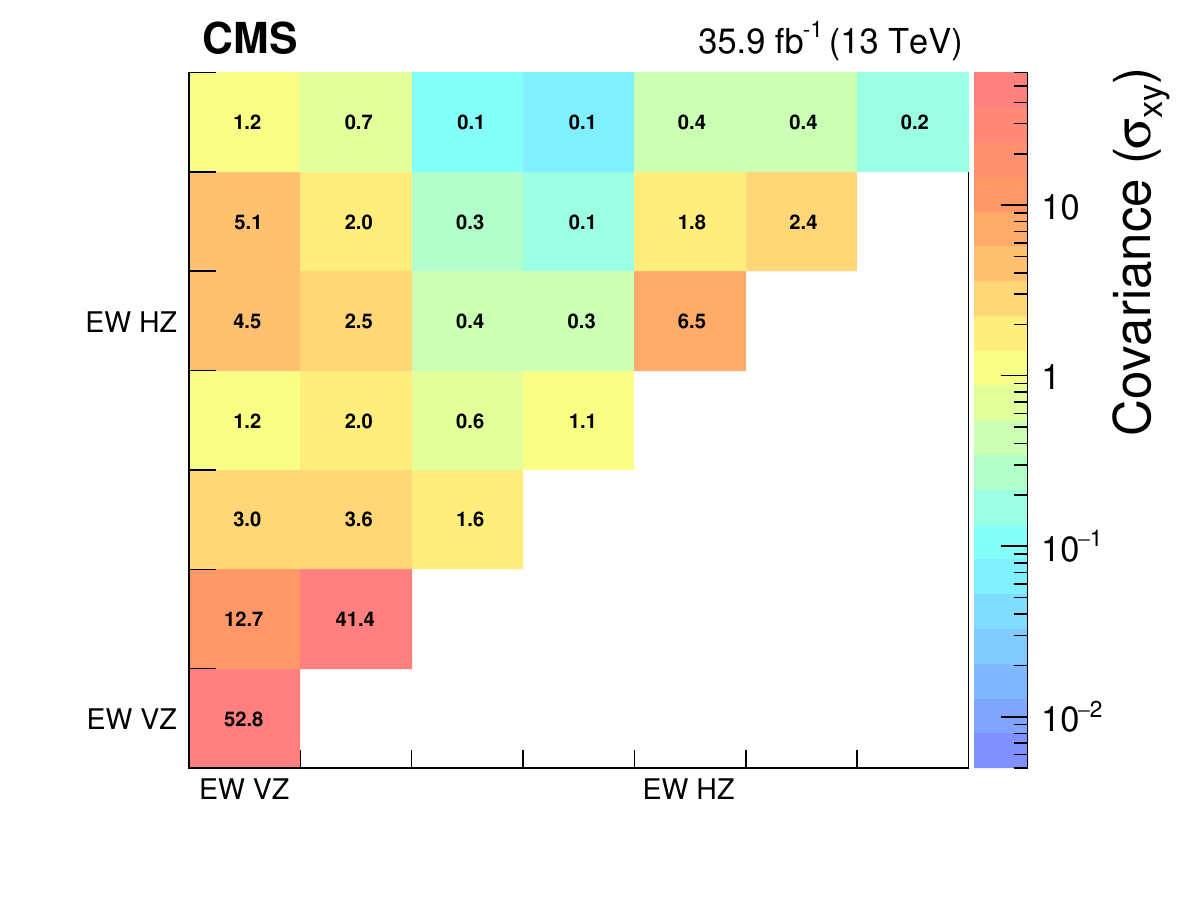}
\includegraphics[width=0.8\linewidth]{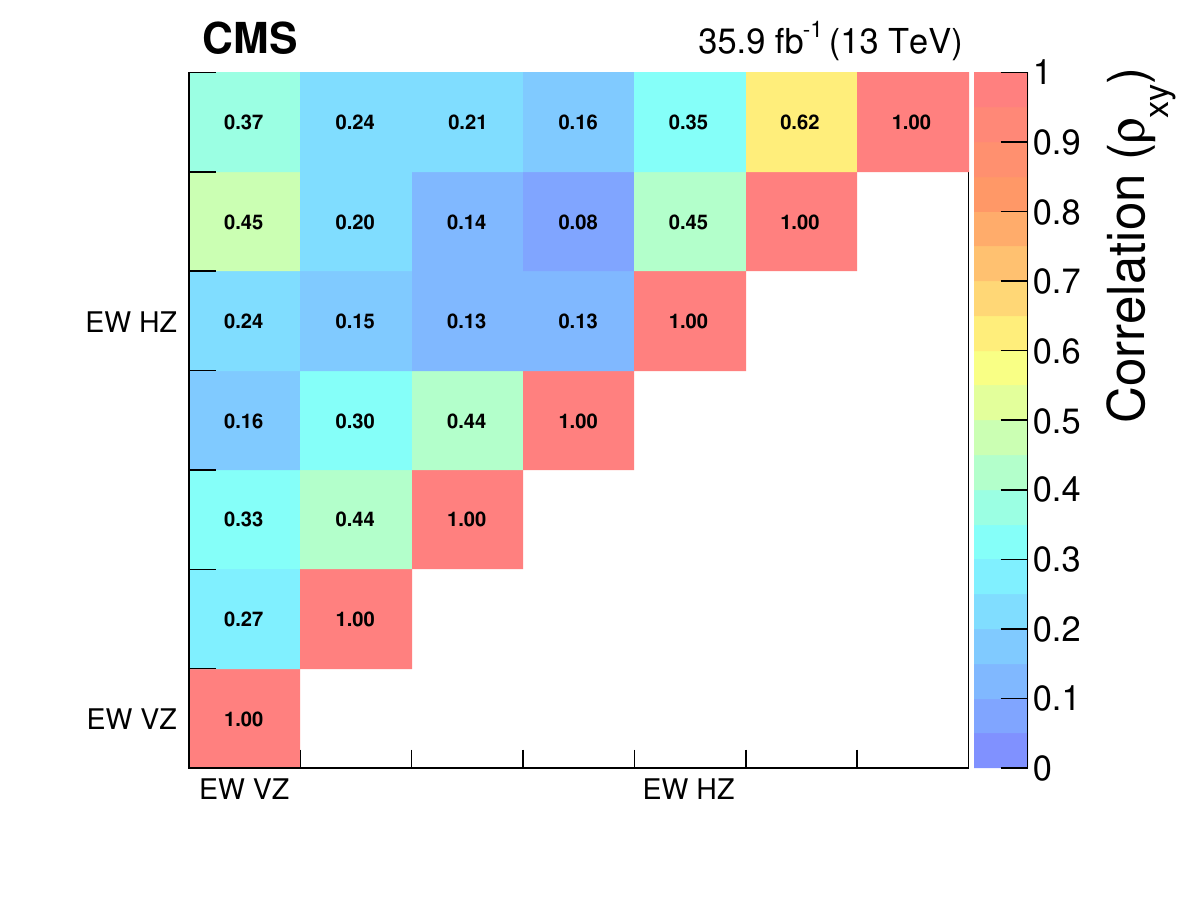}
\caption{\label{fig:correlationEWK}
The covariance (upper) and correlation (lower) matrices for the background predictions in the on-\PZ EW-production SRs.
Within each SR, the individual \ptmiss bins are shown in increasing order starting from 100\GeV.
The matrices are symmetric, but only the entries along and above the diagonal are shown for simplicity.
}
\end{figure}

\begin{figure}[htbp]
\centering
\includegraphics[width=0.8\linewidth]{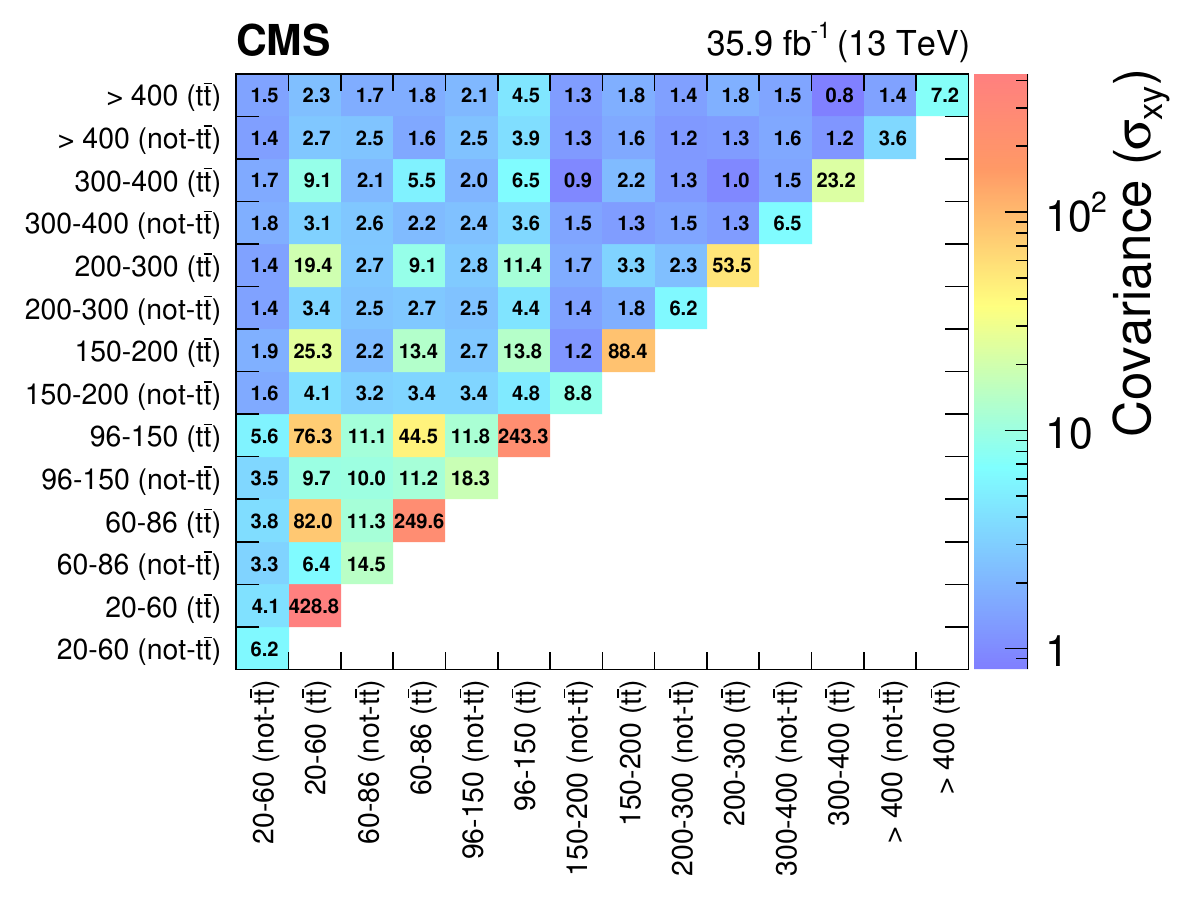}
\includegraphics[width=0.8\linewidth]{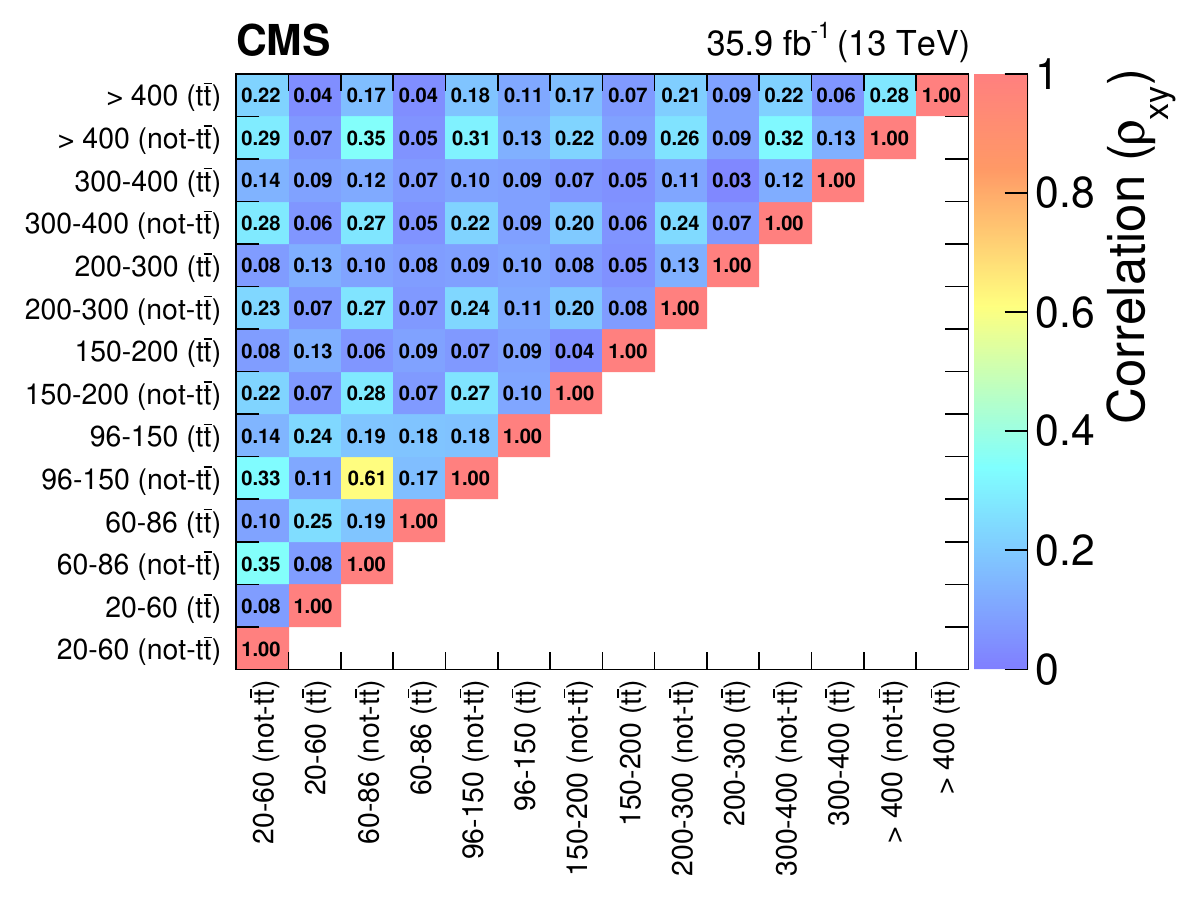}
\caption{\label{fig:correlationEdge}
The covariance (upper) and correlation (lower) matrices for the background predictions in the edge strong-production SRs.
The matrices are symmetric, but only the entries along and above the diagonal are shown for simplicity.
}
\end{figure}

In order to facilitate the interpretation of the analysis results in other models, we provide the covariance
and correlation matrices for the background predictions in the different SRs. Figure~\ref{fig:correlationA}
shows a graphical representation of the covariance (upper) and correlation (lower) matrices for the on-\PZ strong-production SRs.
Figure~\ref{fig:correlationEWK} shows
the same matrices for the on-\PZ electroweak-production SRs,
and Fig.~\ref{fig:correlationEdge} shows the corresponding matrices for the edge strong-production SRs.
Because of potential overlaps in selected events, only the strong- or electroweak-production on-\PZ SRs should be used
for interpretation, not both sets simultaneously.
This information can
be used to construct a simplified likelihood for models of new physics, as described in Ref.~\cite{Collaboration:2242860}.

\cleardoublepage \section{The CMS Collaboration \label{app:collab}}\begin{sloppypar}\hyphenpenalty=5000\widowpenalty=500\clubpenalty=5000\textbf{Yerevan Physics Institute,  Yerevan,  Armenia}\\*[0pt]
A.M.~Sirunyan, A.~Tumasyan
\vskip\cmsinstskip
\textbf{Institut f\"{u}r Hochenergiephysik,  Wien,  Austria}\\*[0pt]
W.~Adam, F.~Ambrogi, E.~Asilar, T.~Bergauer, J.~Brandstetter, E.~Brondolin, M.~Dragicevic, J.~Er\"{o}, M.~Flechl, M.~Friedl, R.~Fr\"{u}hwirth\cmsAuthorMark{1}, V.M.~Ghete, J.~Grossmann, J.~Hrubec, M.~Jeitler\cmsAuthorMark{1}, A.~K\"{o}nig, N.~Krammer, I.~Kr\"{a}tschmer, D.~Liko, T.~Madlener, I.~Mikulec, E.~Pree, D.~Rabady, N.~Rad, H.~Rohringer, J.~Schieck\cmsAuthorMark{1}, R.~Sch\"{o}fbeck, M.~Spanring, D.~Spitzbart, W.~Waltenberger, J.~Wittmann, C.-E.~Wulz\cmsAuthorMark{1}, M.~Zarucki
% ...
\vskip\cmsinstskip
% ...
1:~~Also at Vienna University of Technology, Vienna, Austria\\
% ...

\end{sloppypar}
\end{document}